\documentstyle[12pt]{article}

        {\newpage
        \setcounter{page}{1}}

\renewcommand{\title}[1]{%
        {\begin{center}
        \Large\bf #1
        \end{center}}
        \vskip .3in}

\renewcommand{\author}[1]{%
        {\begin{center}
        #1
        \end{center}}}

\renewcommand{\abstract}[1]{%
        \begin{center}%
        {\vspace{1em}\vspace{0pt}\bf Abstract}%
        \end{center}%
        \noindent #1}

\renewcommand{\date}[1]{%
        \begin{center}%
        #1%
        \end{center}}

        {\end{thebibliography}}

\makeatletter
        \@addtoreset{equation}{section}%
\makeatother

\newcommand{\beq}{\begin{eqnarray}}
\newcommand{\eeq}{\end{eqnarray}}

\newcommand{\mybar}[1]%
        {\kern 0.8pt\overline{\kern -0.8pt#1\kern -0.8pt}\kern 0.8pt}
\newcommand{\sla}[1]%
        {\raise.15ex\hbox{$/$}\kern-.57em #1}
\newcommand{\roughly}[1]%
        {\mathrel{\raise.3ex\hbox{$#1$\kern-.75em\lower1ex\hbox{$\sim$}}}}

\newcommand{\drawsquare}[2]{\hbox{%
\rule{#2pt}{#1pt}\hskip-#2pt
\rule{#1pt}{#2pt}\hskip-#1pt
\rule[#1pt]{#1pt}{#2pt}}\rule[#1pt]{#2pt}{#2pt}\hskip-#2pt
\rule{#2pt}{#1pt}}

\newcommand{\Yfund}{\raisebox{-.5pt}{\drawsquare{6.5}{0.4}}}
\newcommand{\Ysymm}{\raisebox{-.5pt}{\drawsquare{6.5}{0.4}}\hskip-0.4pt%
        \raisebox{-.5pt}{\drawsquare{6.5}{0.4}}}
\newcommand{\Ythrees}{\raisebox{-.5pt}{\drawsquare{6.5}{0.4}}\hskip-0.4pt%
          \raisebox{-.5pt}{\drawsquare{6.5}{0.4}}\hskip-0.4pt%
          \raisebox{-.5pt}{\drawsquare{6.5}{0.4}}}
\newcommand{\Yfours}{\raisebox{-.5pt}{\drawsquare{6.5}{0.4}}\hskip-0.4pt%
          \raisebox{-.5pt}{\drawsquare{6.5}{0.4}}\hskip-0.4pt%
          \raisebox{-.5pt}{\drawsquare{6.5}{0.4}}\hskip-0.4pt%
          \raisebox{-.5pt}{\drawsquare{6.5}{0.4}}}
\newcommand{\Yasymm}{\raisebox{-3.5pt}{\drawsquare{6.5}{0.4}}\hskip-6.9pt%
        \raisebox{3pt}{\drawsquare{6.5}{0.4}}}
\newcommand{\Ythreea}{\raisebox{-3.5pt}{\drawsquare{6.5}{0.4}}\hskip-6.9pt%
        \raisebox{3pt}{\drawsquare{6.5}{0.4}}\hskip-6.9pt
        \raisebox{9.5pt}{\drawsquare{6.5}{0.4}}}

\newcommand{\Yadjoint}{\raisebox{-3.5pt}{\drawsquare{6.5}{0.4}}\hskip-6.9pt%
        \raisebox{3pt}{\drawsquare{6.5}{0.4}}\hskip-0.4pt
        \raisebox{3pt}{\drawsquare{6.5}{0.4}}}
\newcommand{\Ysquare}{\raisebox{-3.5pt}{\drawsquare{6.5}{0.4}}\hskip-0.4pt%
        \raisebox{-3.5pt}{\drawsquare{6.5}{0.4}}\hskip-13.4pt%
        \raisebox{3pt}{\drawsquare{6.5}{0.4}}\hskip-0.4pt%
        \raisebox{3pt}{\drawsquare{6.5}{0.4}}}
\newcommand{\Yflavor}{\Yfund + \overline{\Yfund}} 
\newcommand{\Yoneoone}{\raisebox{-3.5pt}{\drawsquare{6.5}{0.4}}\hskip-6.9pt%
        \raisebox{3pt}{\drawsquare{6.5}{0.4}}\hskip-6.9pt%
        \raisebox{9.5pt}{\drawsquare{6.5}{0.4}}\hskip-0.4pt%
        \raisebox{9.5pt}{\drawsquare{6.5}{0.4}}}%

\newcommand{\jref}[4]{{\it #1} {\bf #2}, #3 (#4)}

\newcommand{\MPLA}[3]{\jref{Mod.\ Phys.\ Lett.}{A#1}{#2}{#3}}

\newcommand{\NPB}[3]{\jref{Nucl.\ Phys.}{B#1}{#2}{#3}}

\newcommand{\PLB}[3]{\jref{Phys.\ Lett.}{#1B}{#2}{#3}}
\newcommand{\PR}[3]{\jref{Phys.\ Rep.}{#1}{#2}{#3}}
\newcommand{\PRD}[3]{\jref{Phys.\ Rev.}{D#1}{#2}{#3}}

\newcommand{\PRL}[3]{\jref{Phys.\ Rev.\ Lett.}{#1}{#2}{#3}}

\newcommand{\PTP}[3]{\jref{Prog.\ Theor.\ Phys.}{#1}{#2}{#3}}

\begin{document}

\begin{titlepage}
\begin{center}
{\hbox to\hsize{hep-th/9612207 \hfill  MIT-CTP-2597}}
{\hbox to\hsize{               \hfill  BUHEP-96-46}}

\bigskip
\bigskip
\bigskip
\vskip.2in

{\Large \bf Confinement in $N=1$ SUSY Gauge Theories} \\
\bigskip
{\Large \bf and Model Building Tools} \\

\bigskip
\bigskip
\bigskip
\vskip.2in

{\bf Csaba Cs\'aki$^a$, Martin Schmaltz$^b$ and Witold Skiba$^a$}\\

\vskip.2in

{ \small \it $^a$ Center for Theoretical Physics

Laboratory for Nuclear Science and Department of Physics

Massachusetts Institute of Technology

Cambridge, MA 02139, USA }

\smallskip

{\tt csaki@mit.edu, skiba@mit.edu}

\bigskip
\bigskip


{\small \it $^b$ Department of Physics

Boston University

Boston, MA 02215, USA }

\smallskip

{\tt schmaltz@abel.bu.edu}

\vspace{1.5cm}
{\bf Abstract}\\
\end{center}

\bigskip

We develop a systematic approach to confinement in $N=1$ supersymmetric
theories. We identify simple necessary conditions for theories to
confine without chiral symmetry breaking and to generate a superpotential
non-perturbatively (s-confine). Applying these conditions we identify
all $N=1$ theories with a single gauge group and no tree-level
superpotential which s-confine. We give a complete list of the confined
spectra and superpotentials. Some of these theories are of great
interest for model building. We give several new examples of models
which break supersymmetry dynamically. 

\bigskip

\end{titlepage}

\section{Introduction}
The number of $N=1$ supersymmetric gauge theories for
which we know exact results on their vacuum structure has
been growing steadily in the last two years. The great
progress was sparked by Seiberg's conjectures about the
infrared properties and phase structure of supersymmetric QCD~\cite{Seib}.
Following in his footsteps, others have obtained results on
a whole zoo of theories~[2-13]. Most of the discovered phenomena
follow similar patterns in the different theories, and one is tempted to
ask if there is maybe a more general approach than the
model-specific trial and error procedure that has been
customary thus far.

Whereas a completely general approach that allows one to
understand all the obtained results seems impossibly
difficult to find, we can make much progress by focusing
on the particular phenomenon of confinement. In fact,
a frequently occuring and relatively easily identified
infrared behavior is ``s-confinement". In a previous
publication~\cite{us} we defined an s-confining theory as a theory
for which all the degrees of freedom in the infrared are gauge invariant
composites of the fundamental fields. Furthermore, we demand
that the infrared physics is described by a smooth effective
theory in terms of these gauge invariants. This description
should be valid everywhere on the moduli space of vacua, including the
origin of field space. Finally, we also demand that an s-confining
theory generates a dynamical superpotential.
At the origin of moduli space all global symmetries of
the theory are unbroken and the global anomalies of the microscopic
theory are matched by the macroscopic gauge invariants of
the effective theory.

The best-known example of a theory which has been conjectured to
be s-confining is supersymmetric QCD (SQCD) with $N$ colors
and $F=N+1$ flavors of fundamental and antifundamental matter,
$Q$ and $\bar{Q}$~\cite{Seib,CERN}.
The gauge invariant confined degrees of freedom are
mesons $M=Q\bar{Q}$ and baryons $B=Q^N$, $\bar{B}=\bar{Q}^N$.
At the origin of moduli space, all components of the mesons
and baryons are massless, and they interact via the
confining superpotential
\beq
  W={1 \over \Lambda^{2N-1}}(\det M-B M \bar{B}) .
  \label{SQCDpot}
\eeq
This description is also valid far from the origin of the moduli space where
the large expectation values of the fields completely break the gauge
group. In such a vacuum the theory is in the Higgs phase.
A smooth gauge invariant description of both the
Higgs and confining vacua of the theory can only exist if there is
no phase transition between the two regions in moduli space.
In particular, there should be no gauge invariant order parameter
that distinguishes the two phases.

To understand this in the example of SQCD, note that the
quarks transform in a faithful representation of the gauge group
$SU(N)$. This implies that arbitrary test charges can be screened by
the dynamical quarks because the vacuum can disgorge quark-antiquark
pairs to screen charges transforming in any representation of the
gauge group. Thus a Wilson loop will always obey a perimeter law because
any charges we might want to use to define the Wilson loop can be screened.
Our definition of s-confinement above necessitates that an s-confining theory
is in such a ``screening-confining'' phase.

This situation should be contrasted with $SU(N)$ with only adjoint matter
or $SO(N)$ with vector matter. In both these cases the matter does
not transform in a faithful representation of the gauge group. Now there
are charges that cannot be screened by the dynamical quarks, and a Wilson
loop can serve as gauge invariant order parameter to distinguish the
Higgs and the confining phases. As a result, such theories cannot have
a single smooth description of both the Higgs and confining phases
of the theory, thus they are not s-confining.

In our previous publication~\cite{us}, we identified two criteria which allow
us to decide whether a given theory can be s-confining without having to
know the explicit infrared description. If we limit our attention to
theories with no tree-level superpotential and only one gauge group,
then the symmetries completely determine the form of any non-perturbatively
generated superpotential. Demanding that this superpotential is smooth
everywhere on the moduli space yields the first of our two conditions.
The other condition arises from studying the theory along some
flat direction in which the gauge group
is broken to a subgroup, and the theory may sufficiently simplify
so that we can understand its infrared physics. If we find a result that
cannot be smoothly connected to a confining phase, we know that
the whole theory is not s-confining either. We discuss the arguments
leading to these two conditions in Section 2 of this paper.
In Section 3 we apply our conditions to identify all theories with a single
gauge group and no tree-level superpotential which s-confine. We
give a complete list of the confined spectra and superpotentials
for all s-confining theories with an arbitrary $SU$, $SO$, $Sp$, or
exceptional gauge group.
Using the results for the s-confining theories, we then demonstrate
in Section 4 how one can generate many more exact solutions for
other models by simply integrating out matter form the s-confining
theories. The models which we obtain in this way display interesting
dynamics: confinement with chiral symmetry breaking, non-perturbatively
generated superpotentials which drive the vacuum to infinity, and
confinement with non-interacting composites.

In Section 5 we turn to applications of our results to
model building. We summarize the various known mechanisms of
dynamical supersymmetry breaking and illustrate each of the
mechanisms with a few examples which we construct using our results of
Sections 3 and 4. Finally, we comment on the possibility of
using our models to construct composite models in the conclusions.
We hope that our tables and superpotentials in Sections 3 and 4
together with the explicit
examples of Section 5 will prove to be a valuable resource for model
builders.

\section{Necessary criteria for s-confinement}

In this section we develop two necessary criteria which
allow us to identify all 
s-confining theories with a simple gauge group and no tree-level
superpotential. The first criterion follows from holomorphy of the
dynamically generated superpotential, which can be determined
using the global symmetries of the theory. This criterion allows us to reduce 
the number of theories that are candidates for s-confinement to a
manageable set. Our second criterion follows from explorations of
regions in moduli space which are easier to understand
than the origin. As will be demonstrated in Section 3, these two conditions
combined are sufficient to identify all s-confining theories with a
single gauge group and no tree-level superpotential.

\subsection{The index constraint}
In this subsection, we derive a simple constraint on the matter content of 
s-confining theories which follows from the requirement of holomorphy
of the confining  superpotential. In theories with a simple gauge group
$G$ and no tree-level superpotential, the symmetries are sufficient to
determine the form of any dynamically generated superpotential
completely~\cite{ADS}. A simple way to prove this makes use of
non-anomalous R-symmetries. Define a $U(1)_R$ symmetry
as follows: all chiral superfields, except for one arbitrarily chosen field 
$\phi_i$, are assigned zero R-charge. The charge $q$ of the remaining field
is determined by requiring anomaly
cancelation of the mixed $G^2 U(1)_R$ anomaly
\beq
(q-1)\mu_i - \sum_{j\ne i} \mu_j + \mu_G =  q \mu_i - \sum_{{\rm \ all}\ j} 
\mu_j + \mu_G = 0,
\label{Ranomaly}
\eeq
where $\mu_i$ is the Dynkin index\footnote{We normalize the index of
the fundamental representations of $SU$ and $Sp$ to 1 and of the vector 
of $SO$ to 2. This definition ensures invariance of the index when
decomposing  representations of $SO(2N)$ under the $SU(N)$ subgroup.
This is relevant to the flows discussed in Section 2.2.}
of the gauge representation of
the field $\phi_i$, and $(q-1)$ is the R-charge of its fermion component.
These three terms arise from the contributions of the fermion components of 
$\phi_i$, of all other matter superfields $\phi_j$ with $j\ne i$, and of the
gauge superfields, respectively.  
The $\mu_j$ are the indices of the remaining matter representations,
they are multiplied by the R-charges $-1$ of the fermion components
of $\phi_j$, and finally $\mu_G$ is the index of the adjoint representation 
of $G$ multiplied by the R-charge $+1$ of the gauginos.
R-invariance of the supersymmetric Lagrangian requires the dynamically
generated superpotential to have R-charge two. This uniquely fixes the
dependence of the superpotential on the field $\phi_i$
\beq
  W \propto \left(  {\phi_i}^{\mu_i} \right)^{2 / ( \sum_j \mu_j -\mu_G )}.
\eeq
To determine the functional dependence on the other superfields, we note
that the global symmetries contain a corresponding $U(1)_R$
symmetry for each of the matter superfields, and the superpotential has to have
R-charge two under each such R-symmetry. Finally, the dependence on the
dynamical scale $\Lambda$ can be determined by dimensional analysis
or using an anomalous R-symmetry~\cite{anomalousR}. The result is
\beq
  W \propto \Lambda^3 \left( \prod_i \left({\phi_i \over \Lambda}\right)^{\mu_i} \right)^{2 / ( \sum_j \mu_j -\mu_G )}.
\label{dynpot}
\eeq
There may be several (or no) possible contractions of
gauge indices, thus the superpotential can be a sum of several terms. 
We require the coefficient of this superpotential to be non-vanishing, then
holomorphy at the origin implies that the exponents of all fields $\phi_i$
are positive integers. Strictly speaking, we should require holomorphy in
the confined degrees of freedom which
would imply that the exponents of composites must be positive integers. Since
we do not want to have to determine all gauge invariants for this argument, we
settle for the weaker constraint on exponents of the fundamental fields.
Therefore,%
\footnote{Other solutions exist if all $\mu_i$ have a common divisor $d$,
then for $\sum_j \mu_j -\mu_G=d\ {\rm or}\ 2 d$ the superpotential
Eq.~\ref{mus} may be regular. We will argue at the end of Section 3 that these
solutions generically do not yield s-confining theories. Another
possibility is that the coefficient of the superpotential above vanishes.
There are examples of confining theories with vanishing superpotentials
in the literature~\cite{ISS}.}
$\sum_j \mu_j -\mu_G=1\ {\rm or}\ 2$. However, in our normalization
of the index, anomaly cancelation further constrains this quantity
to be even, thus
\beq
\sum_j \mu_j -\mu_G = 2 .
\label{mus}
\eeq
This formula summarizes our first necessary condition for s-confinement,
which enables us to rule out most theories immediately.
For example, for SQCD we find that the only candidate
is the theory with $F=N+1$.  
Unfortunately, Eq.~\ref{mus} is not a sufficient condition.
An example for a theory which satisfies Eq.~\ref{mus} but does not s-confine
is $SU(N)$ with an adjoint superfield and one flavor. This theory
is easily seen to be in an Abelian Coulomb phase for generic VEVs of the
adjoint scalars and vanishing VEVs for the fundamentals.
In the following section, we derive another necessary criterion which allows 
us to rule out theories that satisfy the ``index-constraint" but do not 
s-confine.

\subsection{Flows and s-confinement}
The second condition is obtained from studying different regions on the moduli
space of the theory under consideration. A generic supersymmetric theory
with vanishing tree-level superpotential has a large moduli space of vacua.
By definition, an s-confining theory has a smooth description in terms of
gauge invariants everywhere on this moduli space. There should be no
singularities in the superpotential or the K\"ahler potential and there
should be no massless gauge bosons anywhere. 

Thus, we can test a given theory for s-confinement by expanding around
points that are far out in moduli space where the theory simplifies.
In the microscopic theory the gauge group gets broken to a subgroup when
we go out in moduli space by giving large ($\langle\phi\rangle \gg \Lambda$)
expectation values to some fields. In this vacuum, the gauge superfields
corresponding to broken symmetry generators get masses through the
super-Higgs mechanism and the remaining matter fields decompose under
the unbroken subgroup. This ``reduced" theory has a smaller gauge group
and may be easier to understand. If the original theory was s-confining
then its confined description should be valid at this point in moduli
space as well. Therefore, the reduced theory is s-confining if the
original theory was. This statement can be applied in two directions.

\noindent Necessary condition: If the reduced theory does not have a smooth
description with only gauge
invariant degrees of freedom, then the original theory cannot be s-confining.
Sufficient condition: If the original theory is known to
be s-confining, then all possible reduced theories (with a remaining
unbroken gauge group) which the original theory flows to
are s-confining also. The confined spectrum and the confining superpotential
of the reduced theories can be obtained by identifying the corresponding
points in moduli space in the confined description of the original theory and
integrating out all massive fields. In practice, this means identifying the
correct gauge invariant fields which have vacuum expectation values
and integrating out fields which now have mass terms in the superpotential
using their equations of motion.  

The reduced theories will always contain some gauge invariant fields in
the high-energy description which originally transformed under the now
broken gauge generators. These fields do not have any interactions and
are irrelevant to the dynamics of the model.
They can be removed from the theory. In the confined description the
fields corresponding to these gauge singlets are only coupled through
superpotential terms which scale to zero when the VEVs are taken to infinity,
or which are irrelevant in the infrared. 

A non-trivial application of the sufficient condition is given by the
flow from $SU(4)$ with an antisymmetric tensor and 4 ``flavors" of
fundamentals and antifundamentals to $Sp(4)$ with 8 fundamentals.
The $SU(4)$ theory is known to s-confine~\cite{sua}. By giving an
expectation value to the antisymmetric tensor
the gauge group is broken to $Sp(4)$. All components of the
antisymmetric tensor field except for
one singlet are ``eaten" by the super-Higgs mechanism, and the 4 flavors of
fundamentals and antifundamentals become 8 fundamentals of $Sp(4)$.
Applying our sufficient criterion, we conclude that the $Sp$ theory is
s-confining as well. Its confined spectrum and superpotential can be
obtained from the spectrum and superpotential of the $SU(4)$ theory.

A non-trivial example of a theory which can be shown not to s-confine is
$SU(4)$ with three antisymmetric tensors and two flavors. This theory
satisfies our index condition, Eq.~\ref{mus}, and is therefore also a
candidate for s-confinement. 
By giving a VEV to an antisymmetric tensor we can flow from this theory
to $Sp(4)$ with two antisymmetric tensors and four fundamentals. VEVs for the
other antisymmetric tensors let us flow further to $SU(2)$ with eight
fundamentals which is known to be at an interacting fixed point in the
infrared. We conclude that the $SU(4)$ with three tensors and $Sp(4)$
with two tensors and all theories that flow to them cannot be s-confining
either. This allows us to rule out the following chain of theories, all
of which are gauge anomaly free and satisfy Eq.~\ref{mus}:
\beq
\begin{array}{ccccccccc}
              SU(7) & \to & SU(6) & \to & SU(5) & \to & SU(4) & \to & Sp(4) \\
        \Ythreea \, \, 2\, \Yfund \, \,  4\, \overline{\Yfund} & &
        \Ythreea  \, \, \Yasymm \, \,  \Yfund \, \, 3\, \overline{\Yfund} & &
        2\, \Yasymm  \, \, \overline{\Yasymm} \, \, \Yfund \, \, 2\, \overline{\Yfund} & &
        3\, \Yasymm \, \,  2\, \Yfund \, \, 2\, \overline{\Yfund} & &
        2\, \Yasymm \, \,  4\, \Yfund
            \end{array} 
\eeq
Note that a VEV for one of the quark flavors of the $SU(4)$ theory lets us
flow to an $SU(3)$ theory with four flavors which is s-confining. We must
therefore be careful: when we find a flow to an s-confining theory,
it does not follow that the original theory is s-confining as well.
The flow is only a necessary condition. However, in all our examples
we find that a theory with a single gauge group and no tree-level
superpotential is s-confining if it is found to flow to s-confining theories 
in all directions of its moduli space.

\section{All s-confining theories}
In this section, we present our results which we obtained using the
two conditions derived in Section 2. We first created a list of
all theories with a single gauge group and matter content satisfying
the index constraint. Then we studied all possible flat
directions of the individual theories and checked if
they only flow to confining theories.
We summarize these results in the first table of each subsection. In
the first column we list all theories satisfying the index constraint.
In the second column we indicate the result of the flows: theories which
can be shown to have a branch with an unbroken Abelian gauge group we
denote with ``Coulomb branch", for theories which can be shown to flow
to a reduced theory with a non-Abelian gauge group which is not s-confining
we indicate the gauge group of the reduced theory and its matter content,
all other theories are s-confining. 

After identifying all s-confining theories in this way, we explicitly
construct the confined spectra for each s-confining theory.
The group theory used to obtain these results can be found
in Refs.~\cite{slansky,georgi,anomalies}. We present our results in
tables where we indicate the matter content of the ultraviolet
theory in the upper part of the table, and the gauge invariant infrared
spectrum in the lower part. The gauge group and the Young tableaux of the
representations of the matter fields are indicated in the first column.
The other groups correspond to the global symmetries of the theory.
In addition to the listed global symmetries, there is also a global
$U(1)$ with a $G^2 U(1)$ anomaly which is broken by instantons.

Finally, we also give the confining superpotentials when they are not
too long. We denote gauge
invariant composites by their constituents in parenthesis. The relative
coefficients of the different terms can be determined by demanding that
the equations of motion following from this superpotential
reproduce the classical
constraints of the ultraviolet theory. This also constitutes an important
consistency check: in the limit of large
generic expectation values for fields, $\langle\phi\rangle \gg \Lambda$, the
ultraviolet theory behaves classically and all its classical constraints
need to be reproduced by the infrared description. Checking that all these
constraints are reproduced and determining the coefficients is a
very tedious exercise which we only performed for some theories.
Since we have not determined the coefficients of the superpotential terms
for several of the s-confining theories, it may turn out that some of the
terms listed in the confining superpotentials have vanishing coefficients.

A more straightforward and also very powerful consistency check
is provided by the 't~Hooft anomaly matching
conditions. We explicitly checked that all global anomalies match between
the microscopic and macroscopic degrees of freedom in every theory.
Other consistency checks which we performed for a subset of the theories
include explorations of the moduli spaces and adding masses for some
matter fields and checking consistency of the results. More details
on these techniques are described in Section 4.

\subsection{The s-confining $SU(N)$ theories}
In this section, we present all s-confining theories based
on $SU(N)$ gauge groups. We normalize the Dynkin index and the 
anomaly coefficient of the fundamental representation to be one. 
With these conventions, the dimension, index and anomaly coefficient 
of the smallest $SU(N)$ representations are listed below.
\[ \begin{array}{|c|c|c|c|} \hline
{\rm Irrep} & {\rm Dim} & \mu & A \\ \hline
\Yfund & N & 1 & 1 \\
{\rm Adj} & N^2-1 & 2N & 0 \\
\Yasymm & \frac{N(N-1)}{2} & N-2 & N-4 \\
\Ysymm & \frac{N(N+1)}{2} & N+2 & N+4 \\
\Ythreea & \frac{N(N-1)(N-2)}{6} & \frac{(N-3)(N-2)}{2} & 
\frac{(N-3)(N-6)}{2} \\
\Ythrees & \frac{N(N+1)(N+2)}{6} & \frac{(N+2)(N+3)}{2} & 
\frac{(N+3)(N+6)}{2} \\
\Yadjoint & \frac{N(N-1)(N+1)}{3} & N^2-3 & N^2-9 \\
\Ysquare & \frac{N^2(N+1)(N-1)}{12} & \frac{N(N-2)(N+2)}{3} &
           \frac{N(N-4)(N+4)}{3} \\
\Yfours & \frac{N(N+1)(N+2)(N+3)}{24} & \frac{(N+2)(N+3)(N+4)}{6} &
          \frac{(N+3)(N+4)(N+8)}{6} \\
\Yoneoone & \frac{N(N+1)(N-1)(N-2)}{8} & \frac{(N-2)(N^2-N-4)}{2} & 
            \frac{(N-4)(N^2-N-8)}{2} \\  \hline \end{array} 
\]

Because the index of a representation of $SU(N)$ grows like $N^{k-1}$
where $k$ is the number of gauge indices,
there are very few anomaly free representations which satisfy Eq.~\ref{mus}.
These representations are listed in Table 1.
In the first column, we indicate the gauge group and the field content
of the theory. In the second column we give the flows which allowed us to rule
out s-confinement for a given theory. For those theories which do
s-confine we then list the spectra and the confining superpotential
in the following tables. For completeness, we also list those
s-confining theories which are already known in the literature.

\begin{table}
\vspace*{-1cm}
\begin{center}
\begin{tabular}{|ll|l|} \hline
$SU(N)$ & $(N+1) (\Yfund + \overline{\Yfund})$ & s-confining \\
$SU(N)$ & $\Yasymm + N\, \overline{\Yfund} + 4\, \Yfund $ & s-confining \\
$SU(N)$ & $\Yasymm + \overline{\Yasymm} + 3 (\Yfund + \overline{\Yfund})$ &
  s-confining \\
$SU(N)$ & Adj  $+\Yfund + \overline{\Yfund}$ & Coulomb branch \\ \hline
$SU(4)$ & Adj $+ \Yasymm $ & Coulomb branch \\
$SU(4)$ & $3\, \Yasymm + 2 (\Yfund + \overline{\Yfund})$ &
   $SU(2)$: $8\, \Yfund$ \\
$SU(4)$ & $ 4\, \Yasymm + \Yfund + \overline{\Yfund}$ &
   $SU(2)$: $\Ysymm + 4\, \Yfund$  \\
$SU(4)$ & $ 5\, \Yasymm $ & Coulomb branch \\
$SU(5)$ & $ 3 (\Yasymm + \overline{\Yfund}) $ & s-confining \\
$SU(5)$ & $ 2\, \Yasymm + 2\, \Yfund + 4\, \overline{\Yfund}$ & s-confining \\
$SU(5)$ & $ 2 (\Yasymm + \overline{\Yasymm})$ & 
   $Sp(4)$: $3\, \Yasymm + 2\, \Yfund$ \\
$SU(5)$ & $2\, \Yasymm + \overline{\Yasymm} + 2\, \overline{\Yfund} +
  \Yfund$ &  $SU(4)$: $3\, \Yasymm + 
  2 (\Yfund + \overline{\Yfund})$ \\
$SU(6)$ & $2\, \Yasymm + 5\, \overline{\Yfund} + \Yfund$ &
  s-confining \\
$SU(6)$ & $ 2\, \Yasymm + \overline{\Yasymm} + 2\, \overline{\Yfund}$ &
   $SU(4)$: $3\, \Yasymm + 2 (\Yfund + \overline{\Yfund})$ \\
$SU(6)$ & $\Ythreea + 4 (\Yfund + \overline{\Yfund})$ & s-confining \\
$SU(6)$ & $\Ythreea + \Yasymm +  3\, \overline{\Yfund} + \Yfund$
  &  $SU(5)$: $2\, \Yasymm +
  \overline{\Yasymm} + 2\, \overline{\Yfund} + \Yfund $ \\
$SU(6)$ & $\Ythreea + \Yasymm + \overline{\Yasymm}$ & 
  $Sp(6)$: $\Ythreea + \Yasymm + \Yfund$ \\
$SU(6)$ & $2\, \Ythreea + \Yfund + \overline{\Yfund}$ &  $SU(5)$:
  $  2 (\Yasymm + \overline{\Yasymm})$ \\
$SU(7)$ & $2 (\Yasymm + 3\, \overline{\Yfund})$ & s-confining \\
$SU(7)$ & $ \Ythreea + 4\, \overline{\Yfund} + 2\, \Yfund$ &
   $SU(6)$: $\Ythreea + \Yasymm + 3\, \overline{\Yfund}+ \Yfund$ \\
$SU(7)$ & $\Ythreea + \overline{\Yasymm} + \Yfund$ & $Sp(6)$: $\Ythreea + 
   \Yasymm + \Yfund$ \\
  \hline \end{tabular}
\end{center}
\caption{All $SU$ theories satisfying $\sum_j \mu_j -\mu_G = 2$.
This list is finite because the indices of higher index tensor
representations grow very rapidly with the size of the gauge group.
We list the gauge group and the field content of the theories in the first 
column. In the second column, we indicate which theories are s-confining.
For the theories which do not s-confine we give the flows to non s-confining
theories or indicate that there is a Coulomb branch on the moduli space.}
\end{table}

\subsubsection{$SU(N)$ with $(N+1)(\protect\Yfund +
               \overline{\protect\Yfund})$ (SUSY QCD) \protect\cite{Seib}}

\[ 
\begin{array}{c|c|cccc}
& SU(N) & SU(N+1) & SU(N+1) & U(1) & U(1)_R \\ \hline
Q & \Yfund & \Yfund & 1 & 1 & \frac{1}{N+1} \\
\bar{Q} & \overline{\Yfund} & 1 & \Yfund & -1 & \frac{1}{N+1} \\ \hline \hline
Q\bar{Q} & & \Yfund & \Yfund & 0 & \frac{2}{N+1} \\
Q^N & & \overline{\Yfund} & 1 & N & \frac{N}{N+1} \\
\bar{Q}^N & & 1 & \overline{\Yfund} & -N & \frac{N}{N+1} \end{array}
\]

\[ W_{dyn}= \frac{1}{\Lambda^{2N-1} }\Big[ (Q\bar{Q})^{N+1}-(Q^N)(Q\bar{Q})
(\bar{Q}^N)\Big] \]

\subsubsection{$SU(2N)$ with $\protect\Yasymm +2N \ \overline{\protect\Yfund}
               + 4\ \protect\Yfund$ \protect\cite{sua}}

\[ 
\begin{array}{c|c|ccccc}
& SU(2N) & SU(2N) & SU(4) & U(1)_1 & U(1)_2 & U(1)_R \\ \hline
A & \Yasymm & 1 & 1 & 0 & 2N+4 & 0 \\
\overline{Q} & \overline{\Yfund} & \Yfund & 1 & 4 & -2N+2 & 0 \\
Q & \Yfund & 1 & \Yfund & -2N & -2N+2 & \frac{1}{2} \\ \hline \hline
Q\overline{Q} & & \Yfund & \Yfund & 4-2N & -4N+4 & \frac{1}{2} \\
A\overline{Q}^2 &  & \Yasymm & 1 & 8 & -2N+8 & 0 \\
A^N & & 1 & 1 & 0 & 2N^2+4N & 0 \\
A^{N-1}Q^2  & & 1 & \Yasymm & -4N & 2N^2-2N & 1 \\
A^{N-2}Q^4 & & 1 & 1 & -8N & 2N^2-8N & 2 \\
\overline{Q}^{2N} & & 1 & 1 & 8N & -4N^2+4N  & 0 \end{array}
\]

\begin{eqnarray}
\hspace*{-1cm} W_{dyn} &=& \frac{1}{\Lambda^{4N-1}} 
\Big[ (A^N)(Q\overline{Q})^4
(A\overline{Q}^2)^{N-2}+(A^{N-1}Q^2)(Q\overline{Q})^2 (A\overline{Q}^2)^{N-1}
+\nonumber \\  \hspace*{-1cm}
&& (A^{N-2}Q^4)(A\overline{Q}^2)^N +(\overline{Q}^{2N}) (A^N)(A^{N-2}Q^4)
+ (\overline{Q}^{2N}) (A^{N-1}Q^2)^2 \Big] \nonumber
\end{eqnarray}

\subsubsection{$SU(2N+1)$ with $\protect\Yasymm +(2N+1) \
               \overline{\protect\Yfund} + 4\ \protect\Yfund$
               \protect\cite{sua}}

\[ 
\begin{array}{c|c|ccccc}
& SU(2N+1) & SU(2N+1) & SU(4) & U(1)_1 & U(1)_2 & U(1)_R \\ \hline
A & \Yasymm & 1 & 1 & 0 & 2N+5 & 0 \\
\overline{Q} & \overline{\Yfund} & \Yfund & 1 & 4 & -2N+1 & 0 \\
Q & \Yfund & 1 & \Yfund & -2N-1 & -2N+1 & \frac{1}{2} \\ \hline \hline
Q\overline{Q} & & \Yfund & \Yfund & 3-2N & -4N+2 & \frac{1}{2} \\
A\overline{Q}^2 &  & \Yasymm & 1 & 8 & -2N+7 & 0 \\
A^{N}Q  & & 1 & \Yfund & -2N-1 & 2N^2+3N+1 & \frac{1}{2} \\
A^{N-1}Q^3 & & 1 & \overline{\Yfund} & -6N-3 & 2N^2-3N-2 & 
\frac{3}{2} \\
\overline{Q}^{2N+1} & & 1 & 1 & 4(2N+1) & -4N^2+1 & 0 \end{array}
\]

\begin{eqnarray*}
 W_{dyn} &=& \frac{1}{\Lambda^{2N}} \Big[ (A^NQ)(Q\overline{Q})^3 
(A\overline{Q}^2)^{N-1}+(A^{N-1}Q^3)(Q\overline{Q})(A\overline{Q}^2)^{N}+
\\ && (\overline{Q}^{2N+1})(A^N Q)(A^{N-1} Q^3)
\Big]
\end{eqnarray*}

\subsubsection{$SU(2N+1)$ with $\protect\Yasymm +\overline{\protect\Yasymm} +
               3(\protect\Yfund +\ \overline{\protect\Yfund})$}

\begin{displaymath}
\begin{array}{c|c|cccccc}
 &SU(2N\! +\! 1)& SU(3) \hspace{-5pt}& SU(3) \hspace{-5pt} &
U(1)_1 \hspace{-5pt}& U(1)_2  \hspace{-5pt} & U(1)_3 \hspace{-5pt}&
U(1)_R \hspace{-5pt}
 \\ \hline
A & \Yasymm & 1 & 1 & 1 & 0 & -3 & 0 \\
\bar{A} & \overline{\Yasymm} & 1 & 1 & -1 & 0 & -3 & 0 \\
Q & \Yfund & \Yfund & 1 & 0 & 1 & 2N-1 & \frac{1}{3} \\
\bar{Q} & \overline{\Yfund} & 1 & \Yfund & 0 & -1  & 2N-1 & \frac{1}{3} \\ 
\hline \hline
M_k=Q (A\bar{A})^k \bar{Q}& & \Yfund & \Yfund & 0 &
     0 &4N-2 -6 k& \frac{2}{3} \\
H_k=\bar{A} (A\bar{A})^k Q^2& & \Yasymm & 1 & -1 &
     2 &4N-5 -6k& \frac{2}{3} \\
\bar{H}_k=A (A\bar{A})^k \bar{Q}^2& & 1 & \Yasymm  & 1 &
    -2 &4N-5 -6k& \frac{2}{3} \\
B_1=A^N Q & & \Yfund & 1 & N & 1 & -N-1 & \frac{1}{3} \\
\bar{B}_1=\bar{A}^N \bar{Q} & & 1 & \Yfund & -N & -1 & -N-1 & \frac{1}{3} \\
B_3=A^{N-1} Q^3 & & 1 & 1 & N-1 & 3 & 3N & 1 \\
\bar{B}_3=\bar{A}^{N-1} \bar{Q}^3 & & 1 & 1 & -N+1 & -3 & 3N & 1 \\
T_m=(A\bar{A})^m & & 1 & 1 & 0 & 0 & -6m & 0 
\end{array}
\end{displaymath}
where $k=0,\ldots,N-1$ and $m=1,\ldots,N$. The number of terms in the
confining superpotential grows quickly with the size of the gauge group. 
Therefore we only present the superpotential
for the $SU(5)$ theory.
\begin{eqnarray*}
W_{dyn} &=& \frac{1}{\Lambda^9} \Big(
     M_0^3 T_1 T_2 + M_1^3 + T_2 B_3 \bar{B}_3 + T_2 H_0 \bar{H}_0 M_0 +
     T_2 M_1 M_0^2 + T_1^3 M_0^3 + \\
 &&  T_1^2 B_3 \bar{B}_3 +T_1^2 H_0 \bar{H}_0 M_0 + T_1^2 M_1 M_0^2 + 
     T_1 B_1 \bar{B}_1 M_0^2 + T_1 H_0 \bar{H}_0 M_1 + \\
 &&  B_1 \bar{B}_1 H_0 \bar{H}_0 + B_1 \bar{B} M_1 M_0+
     H_1 \bar{H}_1 M_0 + H_1 \bar{H}_0 M_0 T_1 + \bar{H}_1 H_0 M_0 T_1 + \\
 &&  \bar{H}_1 \bar{B}_1 B_3 + H_1 B_1 \bar{B}_3 + H_0 B_1 \bar{B}_3 T_1 +
     \bar{H}_0 \bar{B}_1 B_3 T_1 + H_1 \bar{H}_0 M_1 + \bar{H}_1 H_0 M_1
     \Big)
 \end{eqnarray*}
Note  that the term $T_1 M_1^2 M_0$ is allowed by all symmetries, however
its coefficient is zero, which can be verified by requiring that the
equations of motion reproduce the classical constraints.

\subsubsection{$SU(2N)$ with $\protect\Yasymm +\overline{\protect\Yasymm}
               +3(\protect\Yfund +\ \overline{\protect\Yfund})$}

\begin{displaymath}
\label{su2n}
\begin{array}{c|c|cccccc}
 &  SU(2N) & SU(3) \hspace{-2.2pt} & SU(3)  \hspace{-2.2pt} &
U(1)_1  \hspace{-2pt} & U(1)_2 \hspace{-2pt} & U(1)_3 \hspace{-2pt}
& U(1)_R  \hspace{-2pt}
 \\ \hline
A & \Yasymm & 1 & 1 & 1 & 0 & -3 & 0 \\
\bar{A} & \overline{\Yasymm} & 1 & 1 & -1 & 0 & -3 & 0 \\
Q & \Yfund & \Yfund & 1 & 0 & 1 & 2N-2 & \frac{1}{3} \\
\bar{Q} & \overline{\Yfund} & 1 & \Yfund & 0 & -1  & 2N-2 & \frac{1}{3} \\ 
\hline \hline
M_k=Q (A\bar{A})^k \bar{Q} & & \Yfund & \Yfund & 0 & 0 &
     4N-4 -6 k \hspace{-2pt} & \frac{2}{3} \\
H_m=\bar{A} (A\bar{A})^k Q^2 & & \Yasymm & 1 & -1 & 2 & 
     4N-7 -6m \hspace{-2pt} &  \frac{2}{3} \\
\bar{H}_m=A (A\bar{A})^k \bar{Q}^2 & & 1 & \Yasymm  & 1 & -2 & 
     4N-7 -6m \hspace{-2pt} &  \frac{2}{3} \\
B_0=A^N  & & 1 & 1 & N & 0 & -3N & 0\\
\bar{B}_0=\bar{A}^N & & 1 & 1 & -N & 0 & -3N & 0 \\
B_2=A^{N-1} Q^2 &  &\Yasymm  & 1 & N-1 & 2 & N-1 & \frac{2}{3} \\
\bar{B}_2=\bar{A}^{N-1} \bar{Q}^2 & & 1 & \Yasymm & -N+1 
& -2 &N-1&\frac{2}{3}\\
T_n=(A\bar{A})^n & & 1 & 1 & 0 & 0 & -6n & 0 
\end{array}
\end{displaymath}
where $k=0,\ldots,N-1$, $m=0,\ldots,N-2$ and $n=1,\ldots,N-1$. 
The case of $SU(4)$ is different, because in $SU(4)$ the
two-index antisymmetric tensor is self-conjugate. Therefore there is 
an additional $SU(2)$ global symmetry. The corresponding table is
\begin{displaymath}
\begin{array}{c|c|cccccc}
 & SU(4)&SU(2)&SU(3) & SU(3)& U(1)_1 & U(1)_2 & U(1)_R 
 \\ \hline
A & \Yasymm & \Yfund & 1 & 1 & 0 & -3 & 0 \\
Q & \Yfund & 1 & \Yfund & 1 & 1& 2 & \frac{1}{3} \\
\bar{Q} & \overline{\Yfund} & 1 & 1 & \Yfund & -1  & 2& \frac{1}{3} \\ \hline 
\hline
M_0=Q \bar{Q} & & 1 & \Yfund & \Yfund & 0 & 4 & \frac{2}{3} \\
M_2=QA^2\bar{Q} & & 1 & \Yfund & \Yfund & 0 & -2 & \frac{2}{3} \\
H=AQ^2 & & \Yfund & \overline{\Yfund} & 1 & 2 & 1&  \frac{2}{3} \\
\bar{H}=A\bar{Q}^2 & & \Yfund &1& \overline{\Yfund} & -2 & 1&  \frac{2}{3} \\
T=A^2 & & \Ysymm & 1 & 1 & 0 & -6 & 0 
\end{array}
\end{displaymath}
The superpotential for the $SU(4)$ theory is
\begin{eqnarray*}
W_{dyn}=\frac{1}{\Lambda^7}\Big( T^2M_0^3-12TH\bar{H}M_0-24M_0M_2^2-
24H\bar{H}M_2\Big),
\end{eqnarray*}
where the relative coefficients are fixed by requiring that the equations of 
motion reproduce the classical constraints.

\subsubsection{$SU(6)$ with $\protect\Ythreea +4(\protect\Yfund +
               \overline{\protect\Yfund})$}

\begin{displaymath}
\label{su6}
\begin{array}{c|c|ccccc} 
 & SU(6) & SU(4)& SU(4)& U(1)_1 & U(1)_2 & U(1)_R \\ \hline  
 A & \Ythreea & 1 & 1 & 0 & -4 & -1 \\
 Q & \Yfund & \Yfund & 1 & 1 & 3 & 1 \\
 \bar{Q} & \overline{\Yfund} & 1 & \Yfund & -1 & 3 & 1 \\ \hline \hline
M_0=Q\bar{Q}& & \Yfund & \Yfund & 0 & 6 & 2 \\
M_2=QA^2\bar{Q} & & \Yfund & \Yfund & 0 & -2 & 0 \\
B_1=AQ^3 & & \overline{\Yfund} & 1 & 3 & 5 & 2 \\
\bar{B}_1 =A \bar{Q}^3 & & 1 & \overline{\Yfund} & -3 & 5 & 2 \\
B_3=A^3 Q^3 & & \overline{\Yfund} & 1 & 3 & -3 & 0 \\
\bar{B}_3=A^3 \bar{Q}^3 & & 1 & \overline{\Yfund} & -3 & -3 & 0 \\
T=A^4 & & 1 & 1 & 0 & -16 & 4 \end{array}
\end{displaymath}
\begin{eqnarray*}
\label{su6conf}
W_{dyn}&=&\frac{1}{\Lambda^{11}} \Big( M_0 B_1\bar{B}_1T +B_3\bar{B}_3 M_0
+ M_2^3M_0+TM_2M_0^3+ \\
&& \bar{B}_1B_3M_2+B_1\bar{B}_3M_2 \Big), 
\end{eqnarray*}

\subsubsection{$SU(5)$ with $3(\protect\Yasymm +\overline{\protect\Yfund})$}

\begin{displaymath}
\begin{array}{c|c|cccc}
& SU(5) & SU(3) & SU(3) & U(1) & U(1)_R \\ \hline
A & \Yasymm  &  \Yfund  & 1 & 1 & 0 \\
\bar{Q} & \overline{\Yfund} & 1 & \Yfund & -3 & \frac{2}{3} 
\\ \hline \hline
A\bar{Q}^2 & & \Yfund & \overline{\Yfund} & -5 & \frac{4}{3} \\
A^3\bar{Q} & & \Yadjoint & \Yfund & 0 & \frac{2}{3} \\
A^5 & & \Ysymm & 1 & 5 & 0\end{array}
\end{displaymath}

\[ W_{dyn}=\frac{1}{\Lambda^9} \Big[ (A^5)(A^3\overline{Q})(A\overline{Q}^2)+
(A^3\overline{Q})^3 \Big] \]

\subsubsection{$SU(5)$ with $2\, \protect\Yasymm + 4\
               \overline{\protect\Yfund}+2\, \protect\Yfund$}

\begin{displaymath}
\begin{array}{c|c|cccccc}
& SU(5) & SU(2) & SU(4) & SU(2) & U(1)_1 & U(1)_2 & U(1)_R \\ 
\hline
A & \Yasymm  & \Yfund & 1  & 1 & 0 & -1 & 0\\
\bar{Q} & \overline{\Yfund} & 1 & \Yfund & 1 & 1 & 1 & \frac{1}{3} 
\\ 
Q & {\Yfund} & 1 & 1 & \Yfund & -2 & 1 & \frac{1}{3} \\ \hline \hline
Q\bar{Q} & & 1 & \Yfund & \Yfund & -1 & 2 & \frac{2}{3} \\
A\bar{Q}^2 & & \Yfund & \Yasymm & 1 & 2 & 1 & \frac{2}{3} \\
A^2Q & & \Ysymm & 1 & \Yfund & -2 & -1 & \frac{1}{3} \\
A^3\bar{Q} & & \Yfund  & \Yfund & 1 & 1 & -2 &\frac{1}{3} \\
A^2Q^2\bar{Q} & & 1 & \Yfund & 1 & -3 & 1 & 1 \end{array}
\end{displaymath}

\begin{eqnarray}
 W_{dyn}&=&\frac{1}{\Lambda^9}\Big[ (A^3\overline{Q})^2(Q\overline{Q})^2
+(A^3\overline{Q})(A^2Q^2\overline{Q})(A\overline{Q}^2)\nonumber \\
&+&(A^3\overline{Q})
(A^2Q)(A\overline{Q}^2)(Q\overline{Q}) + (A^2Q)^2(A\overline{Q}^2)^2 \Big] 
\nonumber
\end{eqnarray}

\subsubsection{$SU(6)$ with $ 2\, \protect\Yasymm +5\,  
               \overline{\protect\Yfund} +\protect\Yfund$}

\begin{displaymath}
\begin{array}{c|c|ccccc}
& SU(6) & SU(2) & SU(5) & U(1)_1 & U(1)_2 & U(1)_R \\ 
\hline
A & \Yasymm  & \Yfund & 1  & 0 & 3 & \frac{1}{4}\\
\bar{Q} & \overline{\Yfund} & 1 & \Yfund & 1 & -4 & 0 
\\ 
Q & {\Yfund} & 1 & 1 & -5 & -4 & 0 \\
\hline \hline
Q\bar{Q} & & 1 & \Yfund & -4 & -8 & 0 \\
A\bar{Q}^2 & & \Yfund & \Yasymm & 2 & -5 & \frac{1}{4} \\
A^3 & & \Ythrees & 1 & 0 & 9 & \frac{3}{4} \\
A^3Q\bar{Q}& & \Yfund & \Yfund & -4 & 1 & \frac{3}{4} \\
A^4\bar{Q}^2 & & 1 & \Yasymm & 2 & 4 & 1 \end{array}
\end{displaymath}

\begin{eqnarray}
\hspace*{-1cm} W_{dyn} &=& \frac{1}{\Lambda^{11}} 
\Big[ (A^4\overline{Q}^2)^2(Q\overline{Q})+
(A^4\overline{Q}^2)(A^3Q\overline{Q})(A\overline{Q}^2)+
\nonumber \\
\hspace*{-1cm} &&+
(A^3)(A^3Q\overline{Q})(A\overline{Q}^2)^2+(A^3)^2(A\overline{Q}^2)^2
(Q\overline{Q}) \Big] \nonumber
\end{eqnarray}

Note, that the term $(A^4\overline{Q}^2)(A^3)(A\overline{Q}^2)(Q\overline{Q})$
is allowed by the $U(1)$ symmetries but not by the non-abelian
global symmetries.

\subsubsection{$SU(7)$  with $2\, \protect\Yasymm +6
\, \overline{\protect\Yfund}$}
\label{sec:SU7}

\begin{displaymath}
\begin{array}{c|c|cccc}
 & SU(7) & SU(2) & SU(6) & U(1) & U(1)_R \\ \hline 
A & \Yasymm & \Yfund & 1 & 3 & 0 \\
\bar{Q} & \overline{\Yfund} & 1 & \Yfund & -5 & \frac{1}{3} \\
\hline \hline
H=A\bar{Q}^2 & & \Yfund & \Yasymm & -7 & \frac{2}{3} \\
N=A^4\bar{Q} & & \Ysymm & \Yfund & 7 & \frac{1}{3} \end{array}
\end{displaymath}

\[ W_{dyn}=\frac{1}{\Lambda^{13}} N^2 H^2 
\]

\subsection{The s-confining $Sp(2N)$ theories}

We now discuss the s-confining $Sp(2N)$ 
theories. First, we again summarize the group theoretical properties of the
simplest $Sp(2N)$ representations. Contrary to $SU(N)$ groups there
is no chiral anomaly for $Sp(2N)$ groups. The only requirement on the
field content is that there is no Witten anomaly, this is satisfied
if the sum of the Dynkin indices of the matter fields is even.
$Sp(2N)$ is the subgroup of $SU(2N)$ which leaves the tensor
$J^{\alpha \beta} = ({\bf 1}_{N \times N} \otimes i \sigma_2)^{\alpha \beta}$
invariant. Irreducible tensors of $Sp(2N)$ must be traceless with respect to
$J^{\alpha \beta}$. One can obtain these 
irreducible representations by subtracting traces from the $SU(2N)$ tensors.
The properties of these representations are summarized in the table below.
We use a normalization where the index of the fundamental is one. 
This normalization is consistent with the $Sp(2N)\subset SU(2N)$ embedding,
under which $2N\to 2N$. Thus with these conventions the index
of the matter fields does not change under $SU\to Sp$ decompositions.
The adjoint of $Sp(2N)$ is the two-index symmetric tensor.

\[ \begin{array}{|c|c|c|} \hline
{\rm Irrep} & {\rm Dim} & \mu \\ \hline
\Yfund & 2N & 1 \\
\Yasymm & N(2N-1) -1& 2N-2 \\
\Ysymm & N(2N+1)& 2N+2 \\
\Ythreea & \frac{N(2N-1)(2N-2)}{3}-2N & \frac{(2N-3)(2N-2)}{2}-1 \\
\Ythrees & \frac{N(2N+1)(2N+2)}{3} & \frac{(2N+2)(2N+3)}{2} \\
\Yadjoint & \frac{2N(2N-1)(2N+1)}{3}-2N & (2 N)^2-4 \\ \hline
\end{array}
\]

With this knowledge one can again write down all anomaly-free theories for which
the matter content satisfies Eq.~\ref{mus}. These theories are summarized in
Table 2. In the first column, we indicate the gauge group and
the field content of the theory. The second column gives a possible flow
to a non-s-confining theory or if the theory is s-confining, we state
that in the second column. The only s-confining theories based on $Sp(2N)$ 
groups are the two sequences that are already known in the literature.
We give the spectra and dynamically generated superpotentials of these
theories in the tables below.
\newpage

\begin{table}
\vspace*{-1cm}

\begin{center}
\begin{tabular}{|ll|l|} \hline
$Sp(2N)$ & $(2N+4)\, \Yfund$ & s-confining \\
$Sp(2N)$ & $\Yasymm +6\, \Yfund $ & s-confining \\
$Sp(2N)$ & $\Ysymm +2\, \Yfund $ & Coulomb branch \\ \hline 
$Sp(4)$ & $2\, \Yasymm +4\, \Yfund $& $SU(2)$: $8\, \Yfund$ \\
$Sp(4)$ & $3\, \Yasymm +2\, \Yfund$& $SU(2)$: $\Ysymm +4\, \Yfund$ \\
$Sp(4)$ & $ 4\, \Yasymm$ & $SU(2)$: $2\, \Ysymm$ \\
$Sp(6)$ & $2\, \Yasymm +2\,\Yfund$ & $Sp(4)$: $2\, \Yasymm +4\, \Yfund $ \\
$Sp(6)$ & $\Ythreea +5\,\Yfund$ & $Sp(4)$: $2\, \Yasymm +4\, \Yfund $ \\
$Sp(6)$ & $\Ythreea +\Yasymm +\Yfund$ & $SU(2)$: $\Ysymm +4\, \Yfund$ \\
$Sp(6)$ & $2\, \Ythreea$ & $SU(3)$: $\Ysymm + \overline{\Ysymm}$ \\
$Sp(8)$ & $2\, \Yasymm $& $Sp(4)$: $ 5\,\Yasymm$\\ 
\hline
\end{tabular}
\end{center}
\caption{All $Sp$ theories satisfying $\sum_j \mu_j -\mu_G = 2$.
This list is finite because the indices of higher index tensor
representations grow very rapidly with the size of the gauge group.
We list the gauge group and the field content of the theories in the first 
column. In the second column, we indicate which theories are s-confining.
For the remaining ones we give the flows to non-confining theories
or indicate that there is a Coulomb branch on the moduli space.}
\end{table}

\subsubsection{$Sp(2N)$ with $ (2N+4)\ \protect\Yfund$ \protect\cite{IntPoul}}

\[ \begin{array}{c|c|cc}
& Sp(2N) & SU(2N+4) & U(1)_R \\ \hline
Q & \Yfund & \Yfund & \frac{1}{N+2} \\ \hline \hline
Q^2 & & \Yasymm & \frac{2}{N+2} \end{array}
\]

\[ W_{dyn}=\frac{1}{\Lambda^{2N+1}} (Q^2)^{N+2} \]

\subsubsection{$Sp(2N)$ with $\protect\Yasymm +6\ \protect\Yfund$
               \protect\cite{Cho,oursp}}

\begin{displaymath}
 \begin{array}{c|c|ccc}
    & Sp(2 N) & SU(6) & U(1) & U(1)_R \\ \hline
  A & \Yasymm   & 1     & -3   & 0 \\
  Q & \Yfund    & \Yfund& N-1& \frac{1}{3} \\ \hline \hline
  A^k    & & 1       & -3k  & 0 \\
  QA^mQ   &  & \Yasymm & 2 (N-1) -3 k & \frac{2}{3} 
 \end{array}
\end{displaymath}
Here $k=2,3,\ldots ,N$ and $m=0,1,\ldots ,N-1$. The number of 
terms in the superpotential grows quickly with $N$. For $Sp(4)$ the
superpotential is
\[ W_{dyn}= \frac{1}{\Lambda^5}\Big[ (A^2)(Q^2)^3+(Q^2)(QAQ)^2 \Big]. \]

\subsection{The s-confining $SO(N)$ theories}
$SO(N)$ theories\footnote{We do not distinguish between $SO(N)$ and its 
covering group $Spin(N)$.} are distinct from the $SU$ and $Sp$ theories
because contrary to those groups $SO(N)$ has representations which cannot
be obtained from products of the vector representations. 
These are the spinorial representations. A theory can be s-confining
only if all possible test charges can be screened by the matter
fields. Spinors cannot be screened by matter in the vector
representation of $SO$. Thus, theories without spinorial matter cannot
be s-confining. This restricts the number of possible s-confining
$SO(N)$ theories, because the Dynkin index of the 
spinor representation grows exponentially with the size of the gauge
group. The biggest group for which Eq.~\ref{mus} can be satisfied
with matter including spinor representations is $SO(14)$. 

$SO(N)$ theories (for $N>6$) do not have either chiral or Witten
anomalies. We do not consider the $N\leq 6$ theories because they can
be obtained from our previous results by using the following isomorphisms:
$SO(6)\sim SU(4),\; SO(5)\sim Sp(4),\;
SO(4)\sim SU(2)\times SU(2),\; SO(3)\sim SU(2),\; SO(2)\sim U(1)$.

The spinor representations of $SO(N)$ have different properties depending
on whe\-ther $N$ is even or odd. For odd $N$, there is just one spinor 
representation, while for even $N$ there are two inequivalent spinors.
For $N=4k$ the two spinors are self-conjugate while for $N=4k+2$ the
two spinors are complex conjugate to each other.

We use a normalization where the index of the vector of $SO(N)$ is $2$.
The reason is that under the embedding $SO(2N)\supset SU(N)$ the vector
of $SO(2N)$ decomposes as  $2N\to N+\overline{N}$. If we do not want
the index of the matter fields to change under this decomposition we
need to normalize the index of the vector to two. The fundamental
properties of the smallest $SO(N)$ representations are summarized
in the tables below. The adjoint of $SO(N)$ is the two-index
antisymmetric tensor.

\[ \begin{array}{|c|c|c|} \hline 
\multicolumn{3}{|c|}{SO(2N+1)} \\ \hline
{\rm Irrep} & {\rm Dim} & \mu \\ \hline
\Yfund & 2N+1 & 2 \\
S & 2^N & 2^{N-2} \\
\Yasymm & N(2N+1) & 4N-2 \\
\Ysymm & (N+1)(2N+1)-1& 4N+6 \\ \hline \end{array}
\]
 
\[ \begin{array}{|c|c|c|} \hline 
\multicolumn{3}{|c|}{SO(2N)} \\ \hline
{\rm Irrep} & {\rm Dim} & \mu \\ \hline
\Yfund & 2N & 2 \\
S & 2^{N-1} & 2^{N-3} \\
\bar{S}, (S') &  2^{N-1} & 2^{N-3} \\
\Yasymm & N(2N-1) & 4N-4 \\
\Ysymm & N(2N+1)-1& 4N+4 \\ \hline \end{array}
\]
 
Since the vector and the spinors are the only representations 
that potentially have smaller index than the adjoint, it is clear
that candidates for s-confining theories contain only vectors
and spinors. For odd $N$ we denote the field content by
$(s,v)$, where $s$ is the number of spinors and $v$ is the number of
vectors. For even $N$ we use the notation $(s,s',v)$, where
$s$ and $s'$ are the numbers of matter fields in the two inequivalent
spinor representations and $v$ is the number of vectors.

The $SO(8)$ group requires special attention. The reason is that there
is a group automorphism which permutes the two spinor and the vector
representations. Therefore only relative labelings of the representations 
are meaningful. For example $(4,3,0)$ and $(0,3,4)$ in $SO(8)$ are
equivalent.

With this knowledge of group theory we can write down all theories
which satisfy Eq.~\ref{mus}. These theories are listed in Table 3.
Almost all of these theories are s-confining. The only spectrum that has
been given in the literature~\cite{Pouliot} is for $SO(7)$ with $(5,1)$.
Below we list the spectra and the confining superpotentials for 
the s-confining $SO(N)$ theories. Most of the confining superpotentials
are very complicated. We only list those where the number of
terms in the superpotential is reasonably small.

\begin{table}
\vspace*{-1cm}
\[
\begin{array}{|c|c|l|} \hline
SO(14) & (1,0,5) & \mbox{s-confining} \\
SO(13) & (1,4) & \mbox{s-confining} \\
SO(12) & (1,0,7) & \mbox{s-confining} \\
SO(12) & (2,0,3) & \mbox{s-confining} \\
SO(12) & (1,1,3) & \mbox{s-confining} \\
SO(11) & (1,6) & \mbox{s-confining} \\
SO(11) & (2,2) & \mbox{s-confining} \\
SO(10) & (4,0,1) & \mbox{s-confining} \\
SO(10) & (3,0,3) & \mbox{s-confining} \\
SO(10) & (2,0,5) & \mbox{s-confining} \\
SO(10) & (3,1,1) & \mbox{s-confining} \\
SO(10) & (2,1,3) & \mbox{s-confining} \\
SO(10) & (1,1,5) & \mbox{s-confining} \\
SO(10) & (2,2,1) & \mbox{s-confining} \\
SO(10) & (1,0,7) & SU(4)\;  \mbox{with}\; 3\ \Yasymm +2\ ( \Yflavor ) \\
SO(9) & (4,0) & \mbox{s-confining} \\
SO(9) & (3,2) & \mbox{s-confining} \\
SO(9) & (2,4) & \mbox{s-confining} \\
SO(9) & (1,6) &  SU(4)\;  \mbox{with}\; 3\ \Yasymm +2\ (\Yflavor ) \\
SO(8) & (7,0,0) & \mbox{Coulomb branch}\\
SO(8) & (6,1,0) &  \mbox{Coulomb branch} \\
SO(8) & (5,2,0) &  SU(4)\;  \mbox{with}\; 3\ \Yasymm +2\ (\Yflavor ) \\
SO(8) & (5,1,1) &  SU(4)\;  \mbox{with}\; 3\ \Yasymm +2\ (\Yflavor ) \\
SO(8) & (4,3,0) &  \mbox{s-confining} \\
SO(8) & (4,2,1) &  \mbox{s-confining} \\
SO(8) & (3,3,1) &  \mbox{s-confining} \\
SO(8) & (3,2,2) & \mbox{s-confining} \\
SO(7) & (6,0) & \mbox{s-confining} \\
SO(7) & (5,1) & \mbox{s-confining} \\
SO(7) & (4,2) & \mbox{s-confining} \\
SO(7) & (3,3) & \mbox{s-confining} \\
SO(7) & (2,4) &  SU(4)\;  \mbox{with}\; 3\ \Yasymm +2\ (\Yflavor ) \\
SO(7) & (1,5) &  \mbox{Coulomb branch} \\ \hline
\end{array}
\]
\caption{All $SO(N)$ theories which contain at least one spinor and
satisfy $\sum_j \mu_j -\mu_G = 2$. This list is finite because the
index of the spinor representations grows exponentially with $N$.
We list the gauge group of the theory in the first column and the 
matter content in the second column. As explained in the text, for
odd $N$ $(s,v)$ denotes the number of spinors and the number of
vectors, while for even $N$ $(s,s',v)$ denotes the numbers
of the two inequivalent spinors and vectors. 
In the third column, we indicate which theories are s-confining.
For the remaining ones we give the flows to non-confining theories
or indicate that there is a Coulomb branch on the moduli space.}
\end{table}

\subsubsection{$SO(14)$ with (1,0,5)}

\begin{displaymath}
\begin{array}{c|c|ccc}
& SO(14) & SU(5) & U(1) & U(1)_R \\ \hline
S & 64 & 1 & 5 & \frac{1}{8} \\
Q & \Yfund & \Yfund & -8 & 0 \\ \hline \hline
Q^2 & & \Ysymm & -16 & 0 \\
S^2Q^3 & & \overline{\Yasymm} & -14 & \frac{1}{4} \\
S^4 Q^2 & & \Ysymm & 4 & \frac{1}{2} \\
S^4Q^4 & & \overline{\Yfund} & -12 & \frac{1}{2} \\
S^6Q^3 & & \overline{\Yasymm} & 6 & \frac{3}{4} \\
S^8 & & 1 & 40 & 1 \\
S^8Q^4 & & \overline{\Yfund} & 8 & 1 \end{array}
\end{displaymath}

\begin{eqnarray*}
 \hspace*{-1cm}W_{dyn}&=&\frac{1}{\Lambda^{23}} \Big[ 
(S^8Q^4)^2(Q^2)+  (S^8Q^4)(S^6Q^3)(S^2Q^3)+(S^8Q^4)(S^4Q^4)(S^4Q^2) \\
\hspace*{-1cm} &&
+(S^8)^2(Q^2)^5+ (S^8)(S^6Q^3)(S^2Q^3)(Q^2)^2+(S^4Q^2)^4(Q^2)+
(S^6 Q^3)^2 (S^4 Q^2) (Q^2) \\
\hspace*{-1cm}&&+(S^8)(S^4Q^4)^2(Q^2)+(S^8)(S^4Q^2)^2(Q^2)^3 \\
\hspace*{-1cm}&&
+ (S^6Q^3)(S^2Q^3)(S^4Q^2)^2 + (S^6 Q^3)^2 (S^4 Q^4) \Big]
\end{eqnarray*}
Note that several terms allowed by $U(1)$ symmetries are not allowed
by the full set of global symmetries. For example, the $SU(5)$ contraction
in the term $(S^8Q^4)(S^8)(Q^2)^3$ vanishes, since it is not possible
to make an $SU(5)$ invariant from the third power of a symmetric tensor
and one field in the antifundamental representation. There are more examples
of such terms prohibited by non-abelian global symmetries in other theories
in this section.

\subsubsection{$SO(13)$ with (1,4)}

\[ \begin{array}{c|c|ccc}
& SO(13) & SU(4) & U(1) & U(1)_R \\ \hline
S & 64 & 1 & 1 & \frac{1}{8} \\
Q & \Yfund & \Yfund & -2 & 0 \\ \hline \hline
Q^2 & & \Ysymm & -4 & 0 \\
S^2Q^3 & & \overline{\Yfund} & -4 & \frac{1}{4} \\
S^2Q^2 & & \Yasymm & -2 & \frac{1}{4} \\
S^4Q^4 & & 1 & -4 & \frac{1}{2} \\
S^4Q^3 & & \overline{\Yfund} & -2 & \frac{1}{2} \\
S^4Q^2 & & \Ysymm & 0 & \frac{1}{2} \\
S^4Q & & \Yfund & 2 & \frac{1}{2} \\
S^4 & & 1 & 4 & \frac{1}{2} \\
S^6Q^3 & & \overline{\Yfund} & 0 & \frac{3}{4} \\
S^6Q^2 & & \Yasymm & 2 & \frac{3}{4} \\
S^8Q^3 & & \overline{\Yfund} & 2 & 1 \\
S^8 & & 1 & 8 & 1 \end{array}
\]
Note, that one could add the operator $S^8 Q^4$ to the above list
without affecting anomaly matching. However, there is a mass term
allowed for this operator, and by flowing to this theory from
$SO(14)$ with $(1,0,5)$ one finds that this mass term is generated.
Thus $S^8 Q^4$ is not in the IR spectrum.
Similar operators appear in many other s-confining
$SO(N)$ theories. Since a mass term is always generated for such
operators, we do not include them in any of the forthcoming
s-confining spectra.

\subsubsection{$SO(12)$ with (1,0,7)}

\begin{displaymath}
\begin{array}{c|c|ccc}
& SO(12) & SU(7) & U(1) & U(1)_R \\ \hline
S & 32 & 1 & 7 & \frac{1}{4} \\
Q & \Yfund & \Yfund & -4 & 0 \\ \hline \hline
Q^2 & & \Ysymm & -8 & 0 \\
S^2Q^2 & & \Yasymm & 6 & \frac{1}{2} \\
S^2Q^6 & & \overline{\Yfund} & -10 & \frac{1}{2} \\
S^4 & & 1 & 28 & 1 \\
S^4Q^6 & & \overline{\Yfund} & 4 & 1 \end{array}
\end{displaymath}

\begin{eqnarray*}
 \hspace*{-1cm} W_{dyn}&=&\frac{1}{\Lambda^{19}}\Big[ 
(S^4Q^6)^2(Q^2)+(S^4Q^6)(S^2Q^6)(S^2Q^2) + (S^4) (S^2 Q^2)^2 (Q^2)^5 \\ 
 \hspace*{-1cm}&&+(S^4)(S^2Q^6)^2(Q^2)+(S^2 Q^2)^4 (Q^2)^3 +
(Q^2)^7 (S^4)^2\Big]
\end{eqnarray*}

\subsubsection{$SO(12)$ with (2,0,3)}

\begin{displaymath}
\begin{array}{c|c|cccc}
& SO(12) & SU(2) & SU(3) & U(1) & U(1)_R \\ \hline
S & 32 & \Yfund & 1  & 3 & \frac{1}{8} \\
Q & \Yfund & 1 & \Yfund & -8 & 0 \\ \hline \hline
Q^2 & & 1 & \Ysymm & -16 & 0 \\
S^2 & & 1& 1& 6&\frac{1}{4} \\
S^2Q^2 & & \Ysymm & \overline{\Yfund} & -10 & \frac{1}{4} \\
S^4 & & \Yfours & 1 & 12 & \frac{1}{2} \\
S^4Q^2 & & 1 & \Ysymm &  -4 & \frac{1}{2} \\
{S^4 Q^2}' & & \Ysymm & \overline{\Yfund} & -4 & \frac{1}{2} \\
S^6 & & 1 & 1 & 18 & \frac{3}{4} \\
S^6Q^2 & & \Ysymm & \overline{\Yfund} & 2 & \frac{3}{4} \\
S^8Q^2 & & 1 & \Ysymm & 8 & 1 \end{array}
\end{displaymath}

\subsubsection{$SO(12)$ with (1,1,3)}

\vspace*{0.25cm}

\[ \begin{array}{c|c|cccc}
& SO(12) & SU(3) & U(1)_1 & U(1)_2 & U(1)_R \\ \hline
S & 32 & 1 & 1 & 3 & \frac{1}{8} \\
S' & 32' & 1 & -1 & 3 & \frac{1}{8} \\
Q & \Yfund & \Yfund & 0 & -8 & 0 \\ \hline \hline
Q^2 & & \Ysymm & 0 & -16 & 0 \\
S S' Q^3 & & 1 & 0 & -18 & \frac{1}{4} \\
S^2Q^2 & & \overline{\Yfund} & 2 & -10 & \frac{1}{4} \\
S'^2Q^2 & & \overline{\Yfund} & -2 & -10 & \frac{1}{4} \\
S S' Q & & \Yfund & 0 & -2 & \frac{1}{4} \\
S^4 & & 1 & 4 & 12 & \frac{1}{2} \\
S'^4 & & 1 & -4 & 12 & \frac{1}{2} \\
S^2 S'^2 & & 1 & 0 & 12 & \frac{1}{2} \\
S^3 S' Q^3 & & 1 & 2 & -12 & \frac{1}{2} \\
S'^3SQ^3 & & 1 & -2 & -12 & \frac{1}{2} \\
S^2 S'^2 Q^2 & & \Ysymm & 0 & -4 & \frac{1}{2} \\
{S^2 S'^2 Q^2}' & & \overline{\Yfund} & 0 & -4 & \frac{1}{2} \\
S^3 S' Q & & \Yfund & 2 & 4 & \frac{1}{2} \\
S'^3SQ & & \Yfund & -2 & 4 & \frac{1}{2} \\
S^3 S'^3Q^3 & & 1 & 0 & -6 & \frac{3}{4} \\
S^3 S'^3Q & & \Yfund & 0 & 10 & \frac{3}{4} \\
S^4 S'^2Q^2 & & \overline{\Yfund} & 2 & 2 & \frac{3}{4} \\
S'^4 S^2Q^2 & & \overline{\Yfund} & -2 & 2 & \frac{3}{4} \\
S^4 S'^4 & & 1 & 0 & 24 & 1 \\
S^4 S'^4Q^2 & & \overline{\Yfund} & 0 & 8 & 1 \end{array} \]

\subsubsection{$SO(11)$ with (1,6)}

\[ \begin{array}{c|c|ccc}
& SO(11) & SU(6) & U(1) & U(1)_R \\ \hline
S & 32 & 1 & 3 & \frac{1}{4} \\
Q & \Yfund & \Yfund & -2 & 0 \\ \hline \hline
Q^2 & & \Ysymm & -4 & 0 \\
S^2Q^2 & & \Yasymm & 2 & \frac{1}{2} \\
S^2Q^5 & & \overline{\Yfund} & -4 & \frac{1}{2} \\
S^4 & & 1 & 12 & 1 \\
S^4Q^5 & & \overline{\Yfund} & 2 & 1 \\
S^2Q & & \Yfund & 4 & \frac{1}{2} \\
S^2Q^6 & & 1 & -6 & \frac{1}{2} \end{array} \]

\subsubsection{$SO(11)$ with (2,2)}

\[ \begin{array}{c|c|cccc}
& SO(11) & SU(2) & SU(2) & U(1) & U(1)_R \\ \hline
S & 32 & \Yfund  & 1 & 1 & 0 \\
Q & \Yfund & 1 &  \Yfund & -4 & \frac{1}{2} \\ \hline \hline 
Q^2 & & 1 & \Ysymm & -8 & 1 \\
S^2Q^2 & &\Ysymm & 1 & -6 & 1 \\
S^2Q & & \Ysymm & \Yfund & -2 & \frac{1}{2} \\
S^2 & & 1 & 1 & 2 & 0 \\
S^4 & & \Yfours & 1 & 4 & 0 \\
{S^4}' & & 1 & 1 & 4 & 0 \\
S^4Q^2 & & 1 & \Ysymm & -4 & 1 \\
{S^4 Q^2}' & & \Ysymm & 1 & -4 & 1 \\
S^4Q & & \Ysymm & \Yfund & 0 & \frac{1}{2} \\
S^6Q^2 & & \Ysymm & 1 & -2 & 1 \\
S^6Q & & \Ysymm & \Yfund & 2 & \frac{1}{2} \\
S^8 & & 1 & 1 & 8 & 0 \\
S^8Q & & 1 & \Yfund & 4 & \frac{1}{2} \\
S^4Q & & 1 & \Yfund & 0 & \frac{1}{2} \\
S^6 & & 1 & 1 & 6 & 0 \end{array} \]

\subsubsection{$SO(10)$ with (4,0,1)}

\begin{displaymath}
\begin{array}{c|c|ccc}
& SO(10) & SU(4) & U(1) & U(1)_R \\ \hline
S & 16 & \Yfund & 1  & 0 \\
Q & \Yfund & 1 & -8 & 1 \\ \hline \hline
Q^2 & & 1 & -16 & 2 \\
S^2Q & & \Ysymm & -6 & 1 \\
S^4 & & \Ysquare & 4 & 0 \\
S^6Q & & \Ysymm & -2 & 1 \end{array}
\end{displaymath}

\[ W_{dyn}=\frac{1}{\Lambda^{15}}\Big[ (S^6Q)^2(S^4)+(S^6Q)(S^2Q)(S^4)^2+
(S^2Q)^2(S^4)^3+(S^4)^4(Q^2) \Big] \]

\subsubsection{$SO(10)$ with (3,0,3)}

\begin{displaymath}
\begin{array}{c|c|cccc}
& SO(10) & SU(3) & SU(3) & U(1) & U(1)_R \\ \hline
S & 16 & \Yfund & 1  & 1 & 0 \\
Q & \Yfund & 1 & \Yfund & -2 & \frac{1}{3} \\ \hline \hline
Q^2 & & 1 & \Ysymm & -4 & \frac{2}{3} \\
S^2Q & & \Ysymm & \Yfund & 0  & \frac{1}{3} \\
S^2Q^3 & & \overline{\Yfund} & 1 & -4 & 1 \\
S^4 & & \overline{\Ysymm} & 1 & 4 & 0 \\
S^4Q^2 & & \Yfund & \overline{\Yfund} & 0 & \frac{2}{3}
\end{array}
\end{displaymath}

\begin{eqnarray}
 \hspace*{-1cm} W_{dyn}&=&\frac{1}{\Lambda^{15}} \Big[
(S^4Q^2)^3+(S^4Q^2)^2(S^2Q)^2+(S^4Q^2)^2(S^4)(Q^2)+(S^2Q^3)^2(S^4)^2
\nonumber \\
 \hspace*{-1cm}  &&+(S^2Q)^2(Q^2)^2(S^4)^2 +
(S^2Q)^4(Q^2)(S^4) +(Q^2)^3(S^4)^3 + (S^2 Q)^6 \nonumber \\
 \hspace*{-1cm}  &&+ (S^4) (S^2 Q^3) (S^4 Q^2) (S^2 Q) +
(S^4 Q^2) (S^4) (S^2 Q)^2 (Q^2) \nonumber \\
 \hspace*{-1cm}  && + (S^4 Q^2) (S^4 Q)^4 +
(S^2 Q^3) (S^2 Q)^3 (S^4) \Big] \nonumber
\end{eqnarray}

\subsubsection{$SO(10)$ with (2,0,5)}

\begin{displaymath}
\begin{array}{c|c|cccc}
& SO(10) & SU(2) & SU(5) & U(1) & U(1)_R \\ \hline
S & 16 & \Yfund & 1  & 5 & \frac{1}{4} \\
Q & \Yfund & 1 & \Yfund & -4 & 0 \\ \hline \hline
Q^2 & & 1 & \Ysymm & -8 & 0 \\
S^2Q & & \Ysymm & \Yfund & 6  & \frac{1}{2} \\
S^2Q^3 & & 1& \overline{\Yasymm} & -2 & \frac{1}{2} \\
S^2Q^5 & & \Ysymm & 1 & -10 & \frac{1}{2} \\
S^4 & & 1 & 1 & 20 & 1 \\
S^4Q^4 & & 1 & \overline{\Yfund} & 4 & 1
\end{array}
\end{displaymath}

\subsubsection{$SO(10)$ with (3,1,1)}

\vspace*{0.25cm}
\begin{displaymath}
\begin{array}{c|c|cccc}
& SO(10) & SU(3) & U(1)_1 & U(1)_2 & U(1)_R \\ \hline
S & 16 & \Yfund & 1  &  0  & 0 \\
\bar{S} & \bar{16} & 1 & -3 & 1 & 0 \\
Q & \Yfund & 1 & 0  & -2 & 1 \\ \hline \hline
Q^2 & & 1 & 0 & -4 & 2 \\
S^2Q & & \Ysymm & 2 & -2  & 1 \\
S\bar{S} & & \Yfund & -2 & 1 & 0 \\
S^3\bar{S}Q & & \Yadjoint & 0 & -1 & 1 \\
S^2\bar{S}^2 & & \Ysymm & -4 & 2 & 0 \\
S^4 & & \overline{\Ysymm} & 4 & 0 & 0 \\
S^5\bar{S} & & \Ysymm & 2 & 1 & 0 \\
S^4\bar{S}^2Q & & \Yfund & -2 & 0 & 1 \\
\bar{S}^2 Q & & 1 & -6 & 0 & 1\\
S^3\bar{S}^3Q^2 & & 1 & -6 & -1 & 2
\end{array}
\end{displaymath}

\subsubsection{$SO(10)$ with (2,1,3)}

\vspace*{0.25cm}
\begin{displaymath}
\begin{array}{c|c|ccccc}
& SO(10) & SU(2) & SU(3) & U(1)_1 & U(1)_2 & U(1)_R \\ \hline
S & 16 & \Yfund & 1  & 1 &  1  & 0 \\
\bar{S} & \bar{16} & 1 & 1 & -2 & 1 & \frac{1}{2} \\
Q & \Yfund & 1 & \Yfund & 0  & -2 & 0 \\ \hline \hline
Q^2 & & 1 & \Ysymm & 0 & -4 & 0 \\
S^2Q & & \Ysymm & \Yfund & 2 & 0  & 0 \\
\bar{S}^2Q & & 1 & \Yfund & -4 & 0 & 1 \\
S\bar{S} & & \Yfund & 1 & -1 & 2 & \frac{1}{2} \\
S^2\bar{S}^2 & & \Ysymm & 1 & -2 & 4 & 1 \\
S^2Q^3 &  & 1 & 1 & 2 & -4 & 0 \\
S^3\bar{S}Q & & \Yfund  & \Yfund & 1 & 2  & \frac{1}{2} \\
S^4 & & 1 & 1  & 4 & 4 & 0 \\
S\bar{S}Q^2 & & \Yfund & \overline{\Yfund} & -1 & -2 & 
\frac{1}{2} \\
S^2\bar{S}^2 Q^2 & & 1 & \overline{\Yfund} & -2 & 0 & 1 \\
S^3\bar{S}Q^3 & & \Yfund & 1 & 1 & -2 & \frac{1}{2}
\end{array}
\end{displaymath}

\subsubsection{$SO(10)$ with (1,1,5)}

\[ \begin{array}{c|c|cccc}
& SO(10) & SU(5) & U(1)_1 & U(1)_2 & U(1)_R \\ \hline
S & 16 & 1 & 1 & 5 & \frac{1}{4} \\
\overline{S} & \overline{16} & 1 & -1 & 5 & \frac{1}{4} \\
Q & \Yfund & \Yfund & 0 & -4 & 0 \\ \hline \hline
Q^2 & & \Ysymm & 0 & -8 & 0 \\
S^2Q & & \Yfund & 2 & 6 & \frac{1}{2} \\
\overline{S}^2Q & & \Yfund & -2 & 6 & \frac{1}{2} \\
S\overline{S} & & 1 & 0 & 10 & \frac{1}{2} \\
S^2Q^5 & & 1 & 2 & -10 & \frac{1}{2} \\
\overline{S}^2Q^5 & & 1 & -2 & -10 & \frac{1}{2} \\
S\overline{S}Q^2 & & \Yasymm & 0 & 2 & \frac{1}{2} \\
S\overline{S}Q^4 & & \overline{\Yfund} & 0 & -6 & \frac{1}{2} \\
S^2\overline{S}^2 & & 1 & 0 & 20 & 1 \\
S^2\overline{S}^2Q^4 & & \overline{\Yfund} & 0 & 4 & 1 \end{array} \]

\subsubsection{$SO(10)$ with (2,2,1)}


\[ \begin{array}{c|c|ccccc}
& SO(10) & SU(2) & SU(2) & U(1)_1 & U(1)_2 & U(1)_R \\ \hline
S& 16 & \Yfund & 1 & 1 & 1 & 0 \\
\overline{S} & \overline{16} & 1 & \Yfund & -1 & 1 & 0 \\
Q & \Yfund & 1 & 1 & 0 & -8 & 1 \\ \hline \hline
Q^2 & & 1 & 1 & 0 & -16 & 2 \\
S^2Q & & \Ysymm & 1 & 2 & -6 & 1 \\
\overline{S}^2Q & &1 & \Ysymm & -2 & -6 & 1 \\
S\overline{S} & & \Yfund & \Yfund & 0 & 2 & 0 \\
S^4 & & 1 & 1 & 4 & 4 & 0 \\
\overline{S}^4 & & 1 & 1 & -4 & 4 & 0 \\
S^2\overline{S}^2 & & \Ysymm & \Ysymm & 0 & 4 & 0 \\
S^3\overline{S}Q & & \Yfund & \Yfund & 2 & -4 & 1 \\
\overline{S}^3SQ & & \Yfund & \Yfund & -2 & -4 & 1 \\
S^2\overline{S}^2Q^2 & & 1 & 1 & 0 & -12 & 2 \\
S^4\overline{S}^2Q & & \Ysymm & 1 & 2 & -2 & 1 \\
\overline{S}^4S^2Q & & 1 & \Ysymm & -2 & -2 & 1 \\
S^3\overline{S}^3 & & \Yfund & \Yfund & 0 & 6 & 0 \\
S^6\overline{S}^2 & & 1 & 1 & 4 & 8 & 0 \\
\overline{S}^6S^2 & & 1 & 1 & -4 & 8 & 0 \end{array} \]

\subsubsection{$SO(9)$ with (4,0)}


\[ \begin{array}{c|c|cc}
& SO(9) & SU(4) & U(1)_R \\ \hline
S & 16 & \Yfund & \frac{1}{8} \\ \hline \hline
S^2 & & \Ysymm & \frac{1}{4} \\
S^4 & & \Ysquare & \frac{1}{2} \\
S^6 & & \Ysymm & \frac{3}{4} \end{array} \]

\begin{eqnarray}
 \hspace*{-1cm} W_{dyn}&=&\frac{1}{\Lambda^{13}}\Big[ 
(S^6)^2(S^4)+(S^6)(S^4)^2(S^2) +(S^4)^4+(S^4)^3(S^2)^2 \Big] \nonumber
\end{eqnarray}

\subsubsection{$SO(9)$ with (3,2)}


\[ \begin{array}{c|c|cccc} 
& SO(9) & SU(3) & SU(2) &  U(1) & U(1)_R \\
S & 16 & \Yfund &  1 & 1 & 0 \\
Q & \Yfund & 1 & \Yfund & -3 & \frac{1}{2} \\ \hline \hline
Q^2 & & 1 & \Ysymm & -6 & 1 \\
S^2Q & & \Ysymm & \Yfund & -1 & \frac{1}{2} \\
S^2 & & \Ysymm & 1 & 2 & 0 \\
S^4 & & \overline{\Ysymm} &  1 & 4 & 0 \\
S^2Q^2 & & \overline{\Yfund} & 1 & -4 & 1 \\
S^4Q^2 & & \Yfund & 1 & -2 & 1 \\
S^4Q & & \Yfund & \Yfund & 1 & \frac{1}{2} \end{array} \]

\subsubsection{$SO(9)$ with (2,4)}


\[ \begin{array}{c|c|cccc}
& SO(9) & SU(2) & SU(4) & U(1) & U(1)_R \\ \hline
S & 16 & \Yfund & 1 & 1 & \frac{1}{4} \\
Q & \Yfund & 1 & \Yfund & -1 & 0 \\ \hline \hline
Q^2 & & 1 & \Ysymm & -2 & 0 \\
S^2Q & & \Ysymm & \Yfund & 1 & \frac{1}{2} \\
S^2 & & \Ysymm & 1 & 2 & \frac{1}{2} \\
S^2Q^3 & & 1 & \overline{\Yfund} & -1 & \frac{1}{2} \\
S^2Q^2 & & 1 & \Yasymm & 0 & \frac{1}{2} \\
S^4Q^3 & & 1 & \overline{\Yfund} & 1 & 1 \\
S^2Q^4 & & \Ysymm & 1 & -2& \frac{1}{2} \\
S^4 & & 1 & 1 & 4 & 1 \end{array} \]

\subsubsection{$SO(8)$ with (3,0,4)}

\begin{displaymath}
\begin{array}{c|c|cccc}
 & SO(8) & SU(4) & SU(3) & U(1) & U(1)_R \\
Q & 8_v & \Yfund & 1 & 3 & \frac{1}{4} \\
S & 8_s & 1 & \Yfund & -4 & 0 \\ \hline \hline
Q^2 & & \Ysymm & 1 & 6 & \frac{1}{2} \\
S^2 & & 1 & \Ysymm & -8 & 0 \\
S^2Q^2 & & \Yasymm & \overline{\Yfund} & -2 & \frac{1}{2} \\
S^2Q^4 & & 1 & \Ysymm & 4 & 1 
\end{array}
\end{displaymath}

\begin{eqnarray}
 \hspace*{-1cm} W_{dyn}&=&\frac{1}{\Lambda^{11}} \Big[
(S^2Q^4)^2(S^2)+(S^2Q^4)(S^2Q^2)^2+(S^2Q^2)^3(Q^2)+(S^2)^3(Q^2)^4
\nonumber \\
 \hspace*{-1cm}&& + (S^2Q^2)^2(S^2)(Q^2)^2\Big]
\nonumber \end{eqnarray}

\subsubsection{$SO(8)$ with (2,1,4)}

\begin{displaymath}
\begin{array}{c|c|ccccc}
 & SO(8) & SU(4) & SU(2) & U(1) & U(1) & U(1)_R \\
Q & 8_v & \Yfund & 1 &  1 & 0 & \frac{1}{4} \\
S & 8_s & 1 & \Yfund & -2 & 1 & 0 \\ 
S' & 8_c & 1 & 1 & 0 & -2 & 0 \\ \hline \hline
Q^2 & & \Ysymm & 1 & 2 & 0 & \frac{1}{2} \\
S^2 & & 1 & \Ysymm & -4 & 2  & 0 \\
S'^2 & & 1 & 1 & 0 & -4 & 0 \\
S^2Q^2 & & \Yasymm & 1  & -2 & 2 & \frac{1}{2} \\
S^2Q^4 & & 1 & \Ysymm & 0 & 2 & 1 \\
S'^2Q^4 & & 1 & 1 & 4 & -4 & 1 \\
S S' Q & & \Yfund & \Yfund & -1 & -1 & \frac{1}{4} \\
S S' Q^3 & & \overline{\Yfund} & \Yfund & 1 & -1 & \frac{3}{4}
\end{array}
\end{displaymath}

\subsubsection{$SO(8)$ with (3,3,1)}


\[ \begin{array}{c|c|ccccc}
& SO(8) & SU(3) & SU(3) & U(1)_1 & U(1)_2 & U(1)_R \\ \hline
Q & 8_v & 1 & 1 & 0 & 6 & 1 \\
S & 8_s & \Yfund & 1 & 1 & -1 & 0 \\
S' & 8_c & 1 & \Yfund & -1 & -1 & 0 \\ \hline \hline
Q^2 & & 1 & 1 & 0 & 12 & 2 \\
S^2 & & \Ysymm & 1 & 2 & -2 & 0 \\
S'^2 & & 1& \Ysymm & -2 & -2 & 0 \\
S S' Q & & \Yfund & \Yfund & 0 & 4 & 1 \\
S^3 S' Q & & 1 & \Yfund & 2 & 2 & 1 \\
S'^3 SQ & & \Yfund & 1 & -2 & 2 & 1 \\
S^2 S'^2 & & \overline{\Yfund} & \overline{\Yfund} & 0 & -4 & 0 
\end{array} \]

\subsubsection{$SO(8)$ with (2,2,3)}


\[ \begin{array}{c|c|cccccc}
& SO(8) & SU(3) & SU(2) & SU(2) & U(1)_1 & U(1)_2 & U(1)_R \\ \hline
Q & 8_v & \Yfund  & 1 & 1 & 0 & 4 & 0 \\
S & 8_s & 1 & \Yfund & 1 & 1 & -3 & \frac{1}{4} \\
S' & 8_c & 1 & 1 & \Yfund & -1 & -3 & \frac{1}{4} \\ \hline \hline
Q^2 & &\Ysymm &  1 & 1 & 0 & 8 & 0 \\
S^2 & & 1 & \Ysymm & 1 & 2 & -6 & \frac{1}{2} \\
S'^2 & & 1& 1 & \Ysymm & -2 & -6 & \frac{1}{2} \\
S S' Q & & \Yfund & \Yfund & \Yfund & 0 & -2 & \frac{1}{2} \\
S^2 Q^2 & & \overline{\Yfund} & 1 & 1 & 2 & 2 & \frac{1}{2} \\
S'^2 Q^2 & & \overline{\Yfund} & 1 & 1 & -2 & 2 & \frac{1}{2} \\
S S' Q^3 & & 1 & \Yfund & \Yfund & 0 & 6 & \frac{1}{2} \\
S^2 S'^2 & & 1 & 1 & 1 & 0 & -12 & 1 \\
S^2 S'^2 Q^2 & & \overline{\Yfund} & 1 & 1 & 0 & -4 & 1 
\end{array} \]

\subsubsection{$SO(7)$ with (6,0)}


\[ \begin{array}{c|c|cc}
& SO(7) & SU(6) & U(1)_R \\ \hline
S & 8 & \Yfund & \frac{1}{6} \\ \hline \hline
S^2 & & \Ysymm & \frac{1}{3} \\
S^4 & & \overline{\Yasymm} & \frac{2}{3} \end{array} \]

\[ W_{dyn}=\frac{1}{\Lambda^9}\Big[ 
(S^4)^3+(S^4)^2(S^2)^2+ (S^2)^6 \Big] \]

\subsubsection{$SO(7)$ with (5,1) \protect\cite{Pouliot}}


\[ \begin{array}{c|c|ccc}
& SO(7) & SU(5) & U(1) & U(1)_R \\ \hline
S & 8 & \Yfund & 1 & 0 \\
Q & \Yfund & 1 & -5 & 1 \\ \hline \hline
Q^2 & & 1 & -10 & 2 \\
S^2 & & \Ysymm & 2 & 0 \\
S^4 & & \overline{\Yfund} & 4 & 0 \\
S^2Q & & \Yasymm & -3 & 1 \\
S^4Q & & \overline{\Yfund} & -1 & 1 \end{array} \]

\begin{eqnarray}
&& \hspace*{-1cm} W_{dyn}=\frac{1}{\Lambda^9}\Big[
(S^4Q)^2 (S^2) +(S^4Q)(S^2Q)(S^4) + (S^2Q)^2(S^4)(S^2) \nonumber \\
\hspace*{-1cm} &&+(Q^2)(S^2)(S^4)^2 + (S^2)^5 (Q^2) \Big]\nonumber 
\end{eqnarray}

\subsubsection{$SO(7)$ with (4,2)}


\[ \begin{array}{c|c|cccc}
& SO(7) & SU(4) & SU(2) & U(1) & U(1)_R \\ \hline
S & 8 & \Yfund & 1 & 1 & 0 \\
Q & \Yfund & 1 & \Yfund & -2 & \frac{1}{2} \\ \hline \hline
Q^2 & & 1 & \Ysymm & -4 & 1 \\
S^2 & & \Ysymm & 1 & 2 & 0 \\
S^2 Q & & \Yasymm & \Yfund & 0 & \frac{1}{2} \\
S^2 Q^2 & & \Yasymm & 1 & -2 & 1 \\
S^4 & & 1 & 1 & 4 & 0 \\
S^4 Q & & 1 & \Yfund & 2 & \frac{1}{2} \end{array} \]

\begin{eqnarray}
 \hspace*{-1cm} W_{dyn}&=&\frac{1}{\Lambda^9} \Big[ (S^4Q)^2(Q^2)+
(S^4Q)(S^2Q)(S^2Q^2)+ (S^2 Q)^2 (S^2 Q^2) (S^2) \nonumber \\
 \hspace*{-1cm} &&+(S^4)(S^2Q^2)^2+(S^2 Q)^2 (S^2)^2 (Q^2)+
(S^2Q^2)^2(S^2)^2 + (S^2)^4 (Q^2)^2 \nonumber 
\Big] \nonumber 
\end{eqnarray}

\subsubsection{$SO(7)$ with (3,3)}


\[ \begin{array}{c|c|cccc}
& SO(7) & SU(3) & SU(3) & U(1) & U(1)_R \\ \hline
S & 8 & \Yfund & 1 & 1 & 0 \\
Q & \Yfund & 1 & \Yfund & -1 & \frac{1}{3} \\ \hline \hline
Q^2 & & 1 & \Ysymm & -2 & \frac{2}{3} \\
S^2 & & \Ysymm & 1 & 2 & 0 \\
S^2Q & & \overline{\Yfund} & \Yfund & 1 & \frac{1}{3} \\
S^2Q^2 & & \overline{\Yfund} & \overline{\Yfund} & 0 & \frac{2}{3} \\
S^2Q^3 & & \Ysymm & 1 & -1 & 1 \end{array} \]

\begin{eqnarray}
 \hspace*{-1cm} W_{dyn}&=&\frac{1}{\Lambda^9} \Big[
(S^2Q^3)^2 (S^2) +(S^2Q^3)(S^2Q^2)(S^2Q) + (S^2Q^2)^3 + (S^2)^3(Q^2)^3
\nonumber \\
 \hspace*{-1cm} &&+(S^2Q^2)^2(S^2)(Q^2)+
(S^2Q)^2(S^2)(Q^2)^2 + (S^2 Q)^2 (S^2 Q^2) (Q^2) \Big] \nonumber
\end{eqnarray}

\subsubsection{The $SO(N)$ theories with $\sum \mu_i -\mu_G = 4$}

Our normalization for the indices of $SO$ groups is somewhat
non-standard. It follows from demanding that the index
is invariant under flows from $SO(2N)$ groups to their $SU(N)$
subgroups. In the normalization where the index of the vector
is one rather than two, it is obvious that one can obtain
a superpotential that is regular at the origin for
$\sum \mu_i -\mu_G= 1$ or $2$. In our normalization, this
corresponds to $\sum \mu_i -\mu_G= 2$ or $4$.
We have explicitly checked that none of the $\sum \mu_i -\mu_G= 4$
theories are s-confining by identifying
flows to non-s-confining theories.

The $\sum \mu_i -\mu_G=4$ $SO(N)$ theories are examples of the special
case where the confining superpotential can be holomorphic at the
origin without Eq.~\ref{mus} being satisfied. This can only happen
when $\mu_G$ and all $\mu_i$ have a common divisor.
Just like the previously mentioned $\sum \mu_i -\mu_G=4$ $SO(N)$ theories,
such theories are unlikely to s-confine. The reason is that while
Eq.~\ref{mus} is preserved under most flows along flat directions,
the property that $\mu_G$ and all $\mu_i$ have
a common divisor is not. Thus for most such theories one should be able
to find a flow to a non-s-confining theory. We expect that none of these
``common divisor" theories s-confine.

\subsection{Exceptional groups}

The analysis for exceptional groups $G_2$, $F_4$, $E_6$, $E_7$,
and $E_8$ is surprisingly simple. The s-confined spectrum of a
$G_2$ gauge theory with 5 fundamentals has already been  worked out
in Ref.\ \cite{G2,Pouliot}. The representations of $G_2$ are real, thus the 
invariant tensors include the two index symmetric tensor. Furthermore,
there are two totally antisymmetric tensors  with three and four indices,
respectively. Therefore, the confined spectrum is

\subsubsection{$G_2$ with 5 $\protect\Yfund$ \protect\cite{G2}}

\vspace*{0.25cm}

\[ \begin{array}{c|c|cc}
& G_2 & SU(5) & U(1)_R \\ \hline
Q & 7 & \Yfund & \frac{1}{5} \\ \hline \hline
M=Q^2 & & \Ysymm & \frac{2}{5} \\
A=Q^3 & & \overline{\Yasymm} & \frac{3}{5} \\
B=Q^4 & & \overline{\Yfund} & \frac{4}{5} \end{array} \]

\[ W_{dyn}=\frac{1}{\Lambda^7}\Big[
M^5+M^2A^2+M B^2+A^2 B \Big] \]

\subsubsection{The $F_4$, $E_6$, $E_7$ and $E_8$ theories}
  
Theories based on any of the other exceptional gauge groups can be shown
to flow to theories which are not s-confining. This is derived most
easily by starting with the real group $F_4$. The lowest dimensional
representations of $F_4$ are the 26 dimensional fundamental representation
and the 52 dimensional adjoint.
Since any theory with adjoint matter has a Coulomb branch on its moduli
space, we can restrict our attention to theories with only fundamentals.
By giving an expectation value to a fundamental one can break $F_4$ to its
maximal subgroup $SO(9)$. Under $SO(9)$ the representations decompose
as follows: $26 \to 1 + 9 + 16$ and $52 \to 16 + 36$. The 9, 16, 36 are
the fundamental, spinor, and adjoint of $SO(9)$. When
giving an expectation value to a fundamental of $F_4$, the spinor
component of its $SO(9)$ decomposition is eaten. Thus an $F_4$ theory
with $N_f$ fundamentals flows to an $SO(9)$ theory with $N_f$ fundamentals
and $N_f-1$ spinors. For no $N_f$ is this $SO(9)$ theory s-confining,
therefore no $F_4$ theory s-confines.

Using this result, it is easy to show that none of the groups
$E_6$, $E_7$, and $E_8$ s-confine. The lowest dimensional
representations of $E_6$ are the (complex) fundamental
and the adjoint. By giving an expectation value to a fundamental,
one can flow to $F_4$, whereas expectation values for an adjoint
lead to a Coulomb branch. Thus, $E_6$ theories cannot be s-confining
either.

By giving  an expectation value to a field in the 56 dimensional fundamental
representation of $E_7$ one can flow to $E_6$, while an expectation
value for the adjoint again yields a Coulomb branch. For $E_8$
the lowest dimensional representation is the adjoint, again leading to
a Coulomb branch. Thus none of the $E_{6,7,8}$ groups with arbitrary matter
are s-confining.

\section{Obtaining new models by integrating out matter}
In the previous chapter we obtained a low-energy description for many
theories which satisfy $\sum \mu_i-\mu_G=2$. Since a number of these theories
contain matter in vector-like representations one can easily derive
descriptions for theories with smaller matter content by integrating out
fields. In this way we obtain confining theories with a quantum modified
constraint, theories with dynamically generated superpotentials and theories
with multiple branches.

\subsection{Theories with quantum-deformed moduli spaces}
In these theories a classical constraint of the form $\sum (\Pi_i X_i)=0$
(where $X_i$ are gauge invariant operators) is modified quantum mechanically
to $\sum (\Pi_i X_i)=\Lambda^p \Pi_j X_j$. Here, the $X_j$ are some other
combination of the gauge invariant operators, including the possibility that the
quantum modification is just $\Lambda^p$. The power $p$ must necessarily be
positive to reproduce the correct classical limit. Such a modification of the
classical constraint is only possible in theories where $\sum \mu_i-\mu_G=0$.
To show this, consider assigning R-charge zero to every chiral superfield.
This R-symmetry is anomalous and the anomaly has to be compensated by
assigning R-charge $\sum \mu_i-\mu_G$ to the scale of the gauge group
raised to the power of its one loop $\beta$ function coefficient
$\Lambda^{(3 \mu_G - \sum \mu_i)/2}$~\cite{anomalousR}.
Since the constraints have to respect this R-symmetry one immediately
sees that $\Lambda$ can only appear in a constraint if it has vanishing
R-charge. Therefore, we conclude that only theories with
$\sum \mu_i-\mu_G=0$ may exhibit quantum deformed moduli spaces.

We can find all theories satisfying $\sum \mu_i-\mu_G=0$ by simply leaving out
a flavor from the matter contents listed in Tables 1 and 2 
for $SU$ and $Sp$ theories and by leaving out a vector from Table 3
for $SO$ theories. The theories obtained from the s-confining ones are all
confining with a quantum modified constraint. It follows from
the procedure of integrating out a flavor that the form of the quantum
modified constraint is $\sum (\Pi_i X_i)=\Lambda^{\Sigma \mu_i}$.

In those cases where the
s-confining theory contains several meson type fields (e.g.\ $Q\bar{Q}$,
$Q A^2 \bar{Q}$, etc.) there will be additional constraints which are not
modified quantum mechanically~\cite{Cho,oursp}. All constraints
can be implemented by adding them to the superpotential with Lagrange
multipliers. Here, we list only those $SU$ theories which were not previously
known in the literature. Similar results can be obtained from
the s-confining $SO$ theories. In the case of $SO(N)$ theories there is always
one quantum modified constraint, while the total number of constraints
equals the number of operators containing exactly two vectors $Q$
in a symmetric representation of the $Q$-flavor symmetry.

In the following superpotentials we denote Lagrange multipliers by
Greek letters, the notation for the confined fields is defined in
the corresponding tables in Section 3. 

\subsubsection{$SU(4)$ with $2\, \protect\Yasymm +
               2 (\protect\Yfund + \overline{\protect\Yfund})$}

\begin{displaymath}
W = \lambda \Big( 3T^2M_0^2-12TH\bar{H}-24M_2^2-\Lambda^8\Big)
+ \mu \Big( 2M_0M_2+H\bar{H}\Big)
\end{displaymath}

\subsubsection{$SU(5)$ with
               $\protect\Yasymm + \overline{\protect\Yasymm}+
               2 (\protect\Yfund + \overline{\protect\Yfund})$}
\begin{eqnarray*}
W&=& \lambda \Big( 3M_0^2T_1T_2 + T_2H_0\bar{H}_0+2T_2M_0M_1
+3T_1^3M_0^2+T_1^2H_0\bar{H}_0+ \\
&& 2T_1^2M_0M_1+2T_1B_1\bar{B}_1M_0+
B_1\bar{B}_1M_1+H_1\bar{H}_1+\bar{H}_0H_1T_1+ \\
&& H_0\bar{H}_1T_1-\Lambda^{10}
\Big)+\mu \Big( 3M_1^2+T_2M_0^2+T_1^2M_0^2+T_1H_0\bar{H}_0+\\
&&
B_1\bar{B}_1M_0+H_0\bar{H}_1+\bar{H}_0H_1\Big) \\
\end{eqnarray*}

\subsubsection{$SU(5)$ with
               $2\, \protect\Yasymm + \protect\Yfund +
                3\, \overline{\protect\Yfund}$}
\begin{displaymath}
W=\lambda\Big[ (A^3\bar{Q})^2 (Q\bar{Q}) + (A^3 \bar{Q})(A^2 Q) (A\bar{Q}^2) 
              -\Lambda^{10} \Big]
\end{displaymath}

\subsubsection{$SU(6)$ with
               $2\,  \protect\Yasymm + 4\, \overline{\protect\Yfund}$}
\begin{displaymath}
W=\lambda \Big[ (A^4 \bar{Q}^2)^2 + (A^3)^2 (A\bar{Q}^2)^2 - \Lambda^{12} \Big]
  + \mu \Big[ (A^4 \bar{Q}^2) (A\bar{Q}^2) + (A^3) (A\bar{Q}^2)^2 \Big]
\end{displaymath}

\subsubsection{$SU(6)$ with
               $\protect\Ythreea + 3( \protect\Yfund +
                 \overline{\protect\Yfund})$}
\begin{eqnarray*}
W&=& \lambda \left(B_1 \bar{B}_1 T + B_3 \bar{B}_3 + M_2^3 + 
             T M_2 M_0^2 -
             \Lambda^{12} \right) + \nonumber \\
  &&         \mu \left(M_2^2 M_0 + T M_0^3 +
             \bar{B}_1 B_3 + B_1 \bar{B}_3 \right)
\end{eqnarray*}

We have seen that all s-confining $\sum \mu_i-\mu_G=2$ theories result in
confining $\sum \mu_i-\mu_G=0$ theories with a quantum modified constraint
after integrating out a flavor. This does not imply that
$\sum \mu_i-\mu_G=2$ theories which do not s-confine cannot result
in confining theories with quantum modified constraints after eliminating
one flavor. 

As an example we consider $SU(4)$ with $3\, \Yasymm +\Yfund
+\overline{\Yfund}$. The theory with an additional flavor is not s-confining,
it flows to $SU(2)$ with $8\, \Yfund\,$. Moreover, one can explicitly
construct a dual description for $SU(4)$ with $3\, \Yasymm + 2( 
\Yfund +\overline{\Yfund})$ by noting that this theory is equivalent
to an $SO(6)$ theory with $3 \Yfund + 2 (S+\bar{S})$, where $S$ and $\bar{S}$
denote a spinor and its conjugate. This dual can be obtained from the dual
of $SO(10)$ with one spinor and 7 vectors~\cite{Pouliot}.
The confining description with one less flavor is obtained from the $SO(8)$
theory with a spinor and 5 vectors~\cite{Pouliot}. The confining
spectrum is given in the table below.

\begin{displaymath}
\begin{array}{c|c|cccc}
   & SU(4) & SU(3) & U(1)_1 & U(1)_2 & U(1)_R \\ \hline
A  & \Yasymm & \Yfund & 0 & 1 & 0 \\
Q  & \Yfund & 1 & 1 & -3 & 0 \\
\bar{Q} & \overline{\Yfund} & 1 & -1 & -3 & 0 \\ \hline \hline
A^2 & & \Ysymm & 0 & 2 & 0 \\
Q A^2 \bar{Q} & & \overline{\Yfund} & 0 & -4 & 0 \\
Q \bar{Q} & & 1 & 0 & -6 & 0 \\
A^3 Q^2 & & 1 & 2 & -3 & 0 \\
A^3 \bar{Q}^2 & & 1 & -2 & -3 & 0
\end{array}
\end{displaymath}
The quantum modified constraint is
\begin{displaymath}
W=\lambda \Big[ (Q\bar{Q})^2 (A^2)^3 + (A^2) (Q A^2 \bar{Q})^2 + (A^3 Q^2)^2 
                + (A^3 \bar{Q}^2)^2 - \Lambda^8 (Q\bar{Q}) \Big]
\end{displaymath}
Note that one can eliminate the field $(Q\bar{Q})$ from the theory by solving
the quantum modified constraint. The remaining fields match all anomalies 
of the ultraviolet theory. It would be interesting to determine which of
the remaining $\sum \mu_i - \mu_G=0$ theories are confining with a quantum
modified constraint.

\subsection{Dynamically generated runaway superpotentials}

Starting from the confining theories with a quantum deformed 
moduli space one obtains theories with dynamically generated 
run-away superpotentials
by integrating out more flavors. Here we only list the dynamical
superpotentials which one finds by starting with the s-confining
$SU$ theories and which are not already in the literature. 
It is straightforward to obtain similar results from the
s-confining $SO$ theories by integrating out vectors.
Our notation for the composites in the following superpotentials
is defined in the corresponding tables in Section 3.

\subsubsection{$SU(4)$ with $2\, \protect\Yasymm +
               F (\protect\Yfund + \overline{\protect\Yfund})$}

\begin{eqnarray*}
W_{F=1}&=&\frac{\Lambda^9M_0}{6T^2M_0^2+48M_2^2}, \\
W_{F=0}&=&0 \; \; {\rm or}\; \; W_{F=0}=\frac{\Lambda^5}{\sqrt{T^2}}.
\end{eqnarray*}

\subsubsection{$SU(5)$ with
               $\protect\Yasymm + \overline{\protect\Yasymm}+
               F (\protect\Yfund + \overline{\protect\Yfund})$}

\begin{eqnarray*}
W_{F=1}&=&\Big( \Lambda^{11} M_1 \Big) \Big/ \Big(M_1M_0T_1T_2+T_2M_1^2+
 T_1^3M_0M_1 +T_1^2M_1^2+\\ && T_1B_1\bar{B}_1M_1
- (T_2M_0+T_1^2M_0+B_1\bar{B}_1)^2\Big), \\
W_{F=0}&=&\pm \frac{\Lambda^6}{\sqrt{(T_2+T_1^2)(T_1\pm \sqrt{T_2+T_1^2})}}.
\end{eqnarray*}
 
\subsubsection{$SU(5)$ with
               $2\, \protect\Yasymm + 2\, \overline{\protect\Yfund}$ 
\protect\cite{ADSprl}}
\begin{displaymath}
W=\frac{\Lambda^{11}}{(A^3 \bar{Q})^2}.
\end{displaymath}

\subsubsection{$SU(6)$ with
               $\protect\Ythreea + F ( \protect\Yfund +
                 \overline{\protect\Yfund})$}
\label{su6dyn}
\begin{eqnarray*}
W_{F=2}&=&\frac{\Lambda^{13} M_2 M_0}{T (M_2 M_0)^2 - (M_2^2 + T M_0^2)^2}, \\
W_{F=1}&=&\pm \frac{\Lambda^3 \sqrt{x_{\pm}}}{
             \frac{x_+^2 x_{\pm} T^2}{M_0^2 z^3 y^2} + \frac{x_{\pm}}{z}}, \\
W_{F=0}&=&0 \; \; {\rm or}\; \; W_{F=0}=\frac{\Lambda^5}{\sqrt{T}},
\end{eqnarray*}
where
\begin{eqnarray*}
 & x= \frac{8 M_2^5}{T} - 10 M_0^2 M_2^3 + 2 M_0^4 M_2 T, \; \;
    y= M_2^2 - M_0^2 T, & \\
 & z= 4 M_2^2 - M_0^2 T, \hspace{0.5cm}  x_{\pm}= x \pm \frac{y z^{3/2}}{T}. &
\end{eqnarray*}

\subsection{Theories with multiple branches}
When integrating out flavors from a few of the s-confining theories we
find that there are multiple possible solutions for the superpotential:
one or more solutions with a dynamically generated term, and a
solution with vanishing superpotential.
This indicates that such theories have several branches
of vacua. There is not only a moduli space with a smooth continuous
parameterization but there is also a discrete parameter distinguishing
a discrete set of vacua. In our examples there are two sets of vacua which
are characterized by $W=0$ with a non-trivial moduli space, and $W\propto
\frac{1}{{\rm fields}}$ without a stable vacuum~\cite{SO,Cho,oursp}.
A consistency check on the assumption that the 
branch with vanishing superpotential describes a confining theory is that
the 't~Hooft anomaly matching conditions are satisfied.
In addition to the previously described $SU(4)$ with $2\, \Yasymm$
$SU(6)$ with $\Ythreea$, also $SO(14)$ with one spinor field
has multiple branches.

\section{Dynamical supersymmetry breaking}
Our new results on the low-energy behavior of supersymmetric
theories can be used to construct new models of dynamical 
supersymmetry breaking. We begin by reviewing the various mechanisms of
dynamical supersymmetry breaking and then present new models which
illustrate these possibilities.

A sufficient set of conditions for dynamical supersymmetry breaking
is that there are no classical flat directions and that there is
a spontaneously broken global symmetry~\cite{ADS}. In a nutshell, the
argument can be summarized as follows. A spontaneously broken global
symmetry implies the presence of a Goldstone boson. Since there are no
non-compact flat directions, there is no massless scalar which could combine
with the Goldstone boson into a supersymmetric multiplet. Therefore,
supersymmetry must be broken.

The spontaneous breaking of a global symmetry can be achieved by one
of three known mechanisms:
by a dynamically generated superpotential~\cite{ADS}, by a quantum modified 
constraint~\cite{IntThomas}, or by confining dynamics~\cite{ISS}.
In the supersymmetry breaking models based on confining dynamics
a suitably chosen tree-level superpotential combines with the
dynamically generated potential to give an effective O'Raifeartaigh
model. We will give examples of all three mechanisms of dynamical
supersymmetry breaking using our new results presented in the 
previous sections. More complicated mechanisms of dynamical 
supersymmetry breaking appear in product group theories, where an
interplay of the strong gauge dynamics and the presence of tree-level
Yukawa couplings results in dynamical supersymmetry 
breaking~\cite{dsb}.

\subsection{Confining dynamics}

The best known example of this type of models is an
$SU(2)$ theory with a field $Q$ in the three-index symmetric 
representation~\cite{ISS}. It has been argued that this $SU(2)$
theory confines without generating a superpotential for the 
confined field $T=Q^4$. In order to lift the only classical flat
direction, the superpotential term $W=\lambda Q^4$ is added. This
tree-level superpotential becomes a linear term after confinement and
breaks supersymmetry. At low energies the theory is effectively
an O'Raifeartaigh model.

A similar example based on confining dynamics can be found using
the s-confining $SU(7)$ theory with $2\, \Yasymm +
6\, \overline{\Yfund}$~\footnote{This model has been obtained independently 
by A.~Nelson and S.~Thomas~\cite{privcomm}.}. The main difference 
compared to the ISS model is that the confining $SU(7)$ gauge group 
generates a superpotential for the confined fields. The 
field content, the confined degrees of freedom and the confining
superpotential for this theory have been described in Section~\ref{sec:SU7}.
In order to lift the flat directions we add
the following renormalizable tree-level superpotential:
\begin{displaymath}
W_{tree}=A^1 \bar{Q}_1 \bar{Q}_2 + A^1 \bar{Q}_3 \bar{Q}_4 +
         A^1 \bar{Q}_5 \bar{Q}_6 + A^2 \bar{Q}_2 \bar{Q}_3 +
         A^2 \bar{Q}_4 \bar{Q}_5 + A^2 \bar{Q}_6 \bar{Q}_1. 
\end{displaymath}
A detailed analysis shows that this superpotential lifts all flat directions
but preserves a $U(1)\times U(1)_R$ global symmetry. After confinement of
the $SU(7)$ gauge group the superpotential is
\begin{displaymath}
W=H^1_{12}+H^1_{34}+H^1_{56}+H^2_{23}+H^2_{45}+H^2_{61} + 
  \frac{1}{\Lambda^{13}} H^2 N^2.
\end{displaymath}
The equations of motion with respect to the fields $H^1_{12}$, $H^1_{34}$,
$H^1_{56}$, $H^2_{23}$, $H^2_{45}$ and $H^2_{61}$ force non-zero
VEVs for some of the $H$ and $N$ fields. This results in spontaneous
breaking of at least one of the global $U(1)$'s. Therefore, supersymmetry 
must be broken as well.

\subsection{Models with a quantum deformed moduli space}
A well-known model of dynamical supersymmetry breaking
based on a theory with a quantum deformed moduli space is the $SU(2)$
theory with four doublets $Q_i$ and six singlets $S^{ij}$~\cite{IntThomas}.
This supersymmetry breaking theory also has a tree-level superpotential
$W=\lambda S^{ij} Q_i Q_j$. Confining $SU(2)$ dynamics results in a
quantum modified constraint ${\rm Pf} M =\Lambda^4$, where $M_{ij}=Q_i Q_j$.
The equations of motion with respect to the singlets $S^{ij}$ give
$M_{ij}=0$. This point is not on the quantum deformed moduli space,
so supersymmetry is broken. 

In this theory the flat directions corresponding to the singlets $S^{ij}$
are not lifted by the tree-level superpotential. After including the
quantum corrections to the K\"ahler potential, the $S^{ij}$ directions
are no longer flat~\cite{Yuri}. This theory is non-chiral, but
it nevertheless breaks supersymmetry. The theory avoids Witten's
no-go theorem for vector-like theories because the Witten index of the theory
changes along the ``pseudo-flat'' direction $S^{ij}~\cite{IntThomas}$.

Similar models can be built using any theory which has a quantum modified
constraint. One can introduce a singlet for every confined degree of freedom
and a tree-level superpotential $W=\sum S^i M_i$. Here, the $S^i$'s are the
singlets and the $M_i$'s are the gauge invariant operators. This superpotential
lifts all flat directions except for the ones corresponding to the gauge singlet
fields. Since the equations of motion with respect to the $S^i$ set the VEVs
of all gauge invariant operators to zero, the quantum modified constraint
cannot be obeyed and supersymmetry is broken. This mechanism can be applied
to any of our theories with quantum deformed moduli space, whether or not
the theory is chiral. 

As an explicit example consider an $SO(7)$ theory with five spinors.
The table of symmetries and invariants is
\begin{displaymath}
\begin{array}{c|c|cc}
   & SO(7) & SU(5) & U(1)_R \\ \hline
S  & 8     & \Yfund & 0 \\ \hline \hline
S^2 &  & \Ysymm & 0 \\
S^4 & & \overline{\Yfund} & 0
\end{array}
\end{displaymath}
The quantum modified constraint is $(S^2)^5 + (S^2) (S^4)^2=\Lambda^{10}$.
We need to introduce the $SO(7)$ gauge singlets $A_{ij}$ and $B^i$,
where $A$ transforms as a conjugate symmetric tensor of $SU(5)$,
while $B$ as a fundamental of $SU(5)$. The superpotential which sets
all $SO(7)$ invariants containing spinors to zero is
\begin{displaymath}
W_{tree}= A_{ij} S^{2,ij} + B^i {S^4}_i.
\end{displaymath}
The full superpotential after confinement is
\begin{displaymath}
W_{tree}= A_{ij} (S^2)^{ij} + B^i (S^4)_i +
\lambda \left[(S^2)^5 + (S^2)(S^4)^2-\Lambda^{10} \right],
\end{displaymath}
where $\lambda$ is a Lagrange multiplier enforcing the constraint.
The equations of motion with respect to the singlets are incompatible
with the quantum modified constraint, hence supersymmetry is broken.

\subsection{Theories with a dynamically generated superpotential}

Most of the examples of models of dynamical supersymmetry breaking
are based on theories with a dynamically generated superpotential.
Here we show two new examples using our theories analyzed in the
previous sections. We can summarize our method for finding these
models as follows: we start with a non-chiral theory which has a 
dynamically generated superpotential. We gauge a global $U(1)$ symmetry
which makes the theory chiral, and we include some singlets which have
non-zero charges under the $U(1)$. The tree-level superpotential 
together with the $U(1)$ D-term lifts all flat directions 
and supersymmetry is seen to be broken after the dynamically
generated superpotential is added.

The first example is based on an $SO(12)\times U(1)$ gauge group 
with matter content

\[
\begin{array}{c|cc}
& SO(12) & U(1) \\
\hline
S & 32 & 1 \\
Q & \Yfund & -4 \\
A & 1 & 8 \\
B & 1 & 2 \\
C & 1 & 6 \end{array}
\]
The independent $SO(12)$ invariant operators and their $U(1)$ charges are

\[ \begin{array}{c|c}
& U(1) \\ \hline
Q^2 & -8 \\
S^4 & 4 \\
A & 8 \\
B & 2 \\
C & 6 \end{array}
\]
The tree-level superpotential 
\[ W_{tree}=AQ^2 \]
sets the $Q^2$ operator to zero. Since the remaining $SO(12)$ 
invariants all have positive $U(1)$ charges, all flat directions are lifted
by the $U(1)$ D-term. The $SO(12)$ gauge group generates a dynamical
superpotential
\[ W_{dyn}=\frac{\Lambda^5}{(Q^2(S^4)^2)^{\frac{1}{5}}}, \]
and the full superpotential is
\[ W=AQ^2 +\frac{\Lambda^5}{(Q^2(S^4)^2)^{\frac{1}{5}}}. \]
The equations of motion can not be satisfied, so we conclude
that this theory breaks supersymmetry. Note that the fields 
$B$ and $C$ are only needed to cancel the $U(1)$ anomalies.

A similar model can be obtained by using the $SU(6)$ theory 
with a three-index antisymmetric tensor. The field content is

\[ \begin{array}{c|cc|c}
& SU(6) & U(1) & SU(3) \\ \hline
A & \Ythreea & 1 & 1 \\
Q & \Yfund & -3 & 1 \\
\overline{Q} & \overline{\Yfund} & -3 & 1 \\
S_1 & 1 & 6 & 1 \\
S_2 & 1 & 4 & 1 \\
S' & 1 & 2 & \Yfund \end{array}
\]
The tree-level superpotential, 
\[ W=S_1 (Q\overline{Q}) + S_2(QA^2\overline{Q}), \]
again lifts all flat directions, and the presence of the dynamically
generated superpotential of Section~\ref{su6dyn} breaks supersymmetry
dynamically.

Clearly, there are other possibilities for constructing similar models.
One can use theories that are chiral without gauging a $U(1)$ symmetry,
such as the $SU(5)$ model with $2 \, \Yasymm$ and $2 \,
\overline{\Yfund}$~\cite{ADSprl}. Or one can make theories
chiral by gauging a larger subgroup of the global symmetries, an example is
the well-known 3-2 model~\cite{ADS}.

\section{Conclusions}
Determining the phase structure of $N=1$ supersymmetric theories with
arbitrary matter content is a very difficult problem. 
We have shown that it is possible to identify all theories which belong to
a certain class of confining theories. A salient feature of these
s-confining theories is that the massless degrees of freedom
are given by the independent gauge invariant chiral operators.
They describe the theory everywhere
on the moduli space including the origin. Another important characteristic
is that there is a non-vanishing superpotential for the confined
degrees of freedom. 

We have given two necessary conditions for a theory to be s-confining.
Using these conditions and the requirement of 't~Hooft anomaly matching
we determined all s-confining theories with a single gauge group. We
listed several new examples of s-confining theories with $SU(N)$
gauge groups. The $SU(N)$ theory with $\Yasymm + \overline{\Yasymm}+
3 (\Yfund + \overline{\Yfund})$ is s-confining for any $N$, while other
new examples s-confine only for particular $N$. There
are no new examples of s-confinement with $Sp(N)$ gauge group.
S-confinement in $SO(N)$ groups requires the presence of at least one
spinorial representation, which restricts $N\leq 14$. It turns out
that most of the $SO(N)$ theories which satisfy our index condition
are s-confining.

The quantity $\sum \mu_i - \mu_G$ which appears in
our index formula is very useful for determining the dynamics of
a given theory. For example, all s-confining theories satisfy
$\sum \mu_i - \mu_G=2$, all theories which confine with a quantum
modified constraint satisfy $\sum \mu_i - \mu_G=0$, and for
$\sum \mu_i - \mu_G=-2$ the dynamically generated superpotential
has the correct $\Lambda$-dependence to be generated by
single instantons.

An interesting possible application of our results on s-confinement
is to composite model building.
Recently, several examples of models with quark-lepton compositeness
have been given~\cite{oursp,NelsonStrassler,Markus}. All these
models rely on the recent exact results for the infrared spectra of
s-confining theories. In these models the dynamically generated
superpotentials can be used to give a natural explanation of the
hierarchy between the top and bottom quark mass~\cite{NelsonStrassler}.
A toy model based on $Sp(6)$ with an antisymmetric tensor~\cite{oursp}
has the interesting feature that it generates three generations of
quarks with a hierarchical structure for the Yukawa couplings dynamically.
We hope that the wealth of new s-confining theories
listed in this paper can be applied to build further interesting
and realistic models of compositeness. 

Our results can also be applied to dynamical supersymmetry breaking.
We have shown several new examples of supersymmetry breaking models
which illustrate different dynamical mechanisms.
These models use either s-confining theories,
or theories obtained from them by integrating out flavors.
Many other new models can be built using our exact results.
\newpage

\section*{Acknowledgments}
We are grateful to Hitoshi Murayama, Lisa Randall and John Terning for useful
conversations. We also thank Peter Cho and Lisa Randall for comments
on the manuscript. C.C. and W.S. are supported in part by the U.S.
Department of Energy under cooperative
agreement \#DE-FC02-94ER40818. M.S. is supported by the U.S.
Department of Energy under grant \#DE-FG02-91ER40676.


\begin{thebibliography}{99}

\bibitem{Seib}
N. Seiberg, \PRD{49}{6857}{1994}, hep-th/9402044;
\NPB{435}{129}{1995}, hep-th/9411149.

\bibitem{anomalousR}
K. Intriligator, R. Leigh, and N. Seiberg, \PRD{50}{1092}{1994}, 
hep-th/9403198.

\bibitem{phases}
K. Intriligator and N. Seiberg, \NPB{431}{551}{1994}, hep-th/9408155.

\bibitem{sua}
H. Murayama, \PLB{355}{187}{1995}, hep-th/9505082;
E. Poppitz and S. Trivedi, \PLB{365}{125}{1996}, hep-th/9507169;
P. Pouliot, \PLB{367}{151}{1996}, hep-th/9510148.

\bibitem{SO}
K. Intriligator and N. Seiberg, \NPB{444}{125}{1995}, hep-th/9503179.

\bibitem{IntPoul}
K. Intriligator and P. Pouliot, \PLB{353}{471}{1995}, hep-th/9505006.

\bibitem{otherexact}
K. Intriligator, R. Leigh, and M. Strassler, \NPB{456}{567}{1995}, hep-th/9506148;
D. Kutasov, \PLB{351}{230}{1995}, hep-th/9503086;
D. Kutasov and A. Schwimmer, \PLB{354}{315}{1995}, hep-th/9505004;
D. Kutasov, A. Schwimmer, and N. Seiberg, \NPB{459}{455}{1996}, hep-th/9510222;
M. Berkooz, \NPB{452}{513}{1995}, hep-th/9505067;
M. Luty, M. Schmaltz and J. Terning, \PRD{54}{7815}{1996}, hep-th/9603034;
N. Evans and M. Schmaltz, hep-th/9609183;
K. Intriligator, \NPB{448}{187}{1995}, hep-th/9505051;
R. Leigh and  M. Strassler, \PLB{356}{492}{1995}, hep-th/9505088;
hep-th/9611020;
J. Brodie and M. Strassler, hep-th/9611197;
O. Aharony, J. Sonnenschein and S. Yankielowicz,  \NPB{449}{509}{1995},
 hep-th/9504113;
O. Aharony, \PLB{351}{220}{1995}, hep-th/9502013.

\bibitem{G2}
I. Pesando, \MPLA{10}{1871}{1995}, hep-th/9506139;
S. Giddings and J. Pierre \PRD{52}{6065}{1995}, hep-th/9506196.

\bibitem{Pouliot}
P. Pouliot, \PLB{359}{108}{1995}, hep-th/9507018;
P. Pouliot and  M. Strassler, \PLB{370}{76}{1996}, hep-th/9510228;
\PLB{375}{175}{1996}, hep-th/9602031.

\bibitem{ISS}
K. Intriligator, N. Seiberg and S. Shenker, \PLB{342}{152}{1995}, 
hep-th/9410203.

\bibitem{Cho}
P. Cho and P. Kraus, \PRD{54}{7640}{1996}, hep-th/9607200.

\bibitem{oursp}
C. Cs\'aki, W. Skiba and M. Schmaltz, hep-th/9607210.

\bibitem{us} C. Cs\'aki, M. Schmaltz and W. Skiba, hep-th/9610139. 

\bibitem{CERN} D. Amati, K. Konishi, Y. Meurice, G. Rossi and G. Veneziano,
\PR{162}{169}{1988} and references therein.

\bibitem{ADS}
I. Affleck, M. Dine, and N. Seiberg, \NPB{256}{557}{1985}. 

\bibitem{slansky}
R. Slansky, Phys. Rept. {\bf 79} (1981) 1.

\bibitem{georgi}
H. Georgi,  ``Lie  Algebras in  Particle Physics. From Isospin to
Unified Theories'' (Frontiers In Physics series, Benjamin/Cummings, 1982).

\bibitem{anomalies} J. Banks and H. Georgi, \PRD{14}{1159}{1976}.

\bibitem{IntThomas}
K. Intriligator and S. Thomas, \NPB{473}{121}{1996}, hep-th/9603158;
K. Izawa and  T. Yanagida, \PTP{95}{829}{1996}, hep-th/9602180.


\bibitem{dsb}
M. Dine, A. Nelson, Y. Nir and Y. Shirman, \PRD{53}{2658}{1996},
hep-ph/9507378;
K. Intriligator and S. Thomas, hep-th/9608046;
E. Poppitz, Y. Shadmi and  S. Trivedi, \NPB{480}{125}{1996},
hep-th/9605113; \PLB{388}{561}{1996}, hep-th/9606184;
C. Cs\'aki, L. Randall and W. Skiba, \NPB{479}{65}{1996}, hep-th/9605108;
C. Cs\'aki, L. Randall, W. Skiba and R. Leigh, \PLB{387}{791}{1996},
hep-th/9607021.

\bibitem{privcomm}
A. Nelson and S. Thomas, private communication.


\bibitem{Yuri}
Y. Shirman, hep-th/9608147.

\bibitem{ADSprl} I. Affleck, M. Dine, and N. Seiberg, \PRL{52}{1677}{1984}.

\bibitem{NelsonStrassler}
M. Strassler, \PLB{376}{119}{1996}, hep-ph/9510342; 
A. Nelson and M. Strassler, hep-ph/9607362;
A. Cohen, D. Kaplan and A. Nelson,  \PLB{388}{588}{1996}, hep-ph/9607394.
 
\bibitem{Markus}
M. Luty, hep-ph/9611387;
M. Luty and R. Mohapatra, hep-ph/9611343.
\end{thebibliography}
\end{document}